\documentclass[12pt, draftclsnofoot, journal, letterpaper,onecolumn]{IEEEtran}

\usepackage[dvips]{color}
\usepackage{epsf}
\usepackage{times}
\usepackage{cite}
\usepackage{epsfig}
\usepackage{graphicx}
\usepackage{amsmath}
\usepackage{amssymb}
\usepackage{amsthm}
\usepackage{amsxtra}
\usepackage{algorithmic}
\usepackage{float}
\usepackage{amsfonts}
\usepackage{dsfont}
\usepackage{subfigure}
\usepackage{url}
\usepackage{epstopdf}

\usepackage[
  open,
  openlevel=2,
  atend
]{bookmark}[2011/04/21]

\newcommand{\AlgorithmCaption}[2]{\textbf{Algorithm {#1}} {#2}} 
\newenvironment{boxedalgorithmic}[2] 
  {\noindent\begin{minipage}[t!]{\columnwidth} \vspace*{0.3cm}\hrule \vspace*{0.1cm}\AlgorithmCaption{#1}{#2} \vspace*{0.1cm}\hrule\footnotesize
   \begin{algorithmic}}
  {\end{algorithmic}\vspace*{0.1cm}\hrule\vspace*{0.3cm}
   \end{minipage}}

\newtheorem{Lemma1}{Lemma}
\newtheorem{Lemma2}[Lemma1]{Lemma}
\newtheorem{Lemma3}[Lemma1]{Lemma}

\newtheorem{Thm1}{Theorem}

\begin{document}

\title{Loss Visibility Optimized Real-time Video Transmission over MIMO Systems}
\author{{\large Amin Abdel Khalek,~\IEEEmembership{Student Member,~IEEE}, \\Constantine Caramanis, ~\IEEEmembership{Member,~IEEE}, \\and Robert W. Heath Jr.,~\IEEEmembership{Fellow,~IEEE}} \\
\thanks{The authors are with the Wireless Networking \& Communications Group in the Department of Electrical and Computer Engineering at UT Austin WNCG, 2501 Speedway Stop C0806, Austin, Texas 78712-1687. Email: \{akhalek,constantine,rheath\}@utexas.edu. This work was supported by the Intel-Cisco Video Aware Wireless Networks (VAWN) Program.}}\maketitle

\vspace{-1cm}

\begin{abstract}


The structured nature of video data motivates introducing video-aware decisions that make use of this structure for improved video transmission over wireless networks. In this paper, we introduce an architecture for real-time video transmission over multiple-input multiple-output (MIMO) wireless communication systems using loss visibility side information. We quantify the perceptual importance of a packet through the packet loss visibility and use the loss visibility distribution to provide a notion of relative packet importance. To jointly achieve video quality and low latency, we define the optimization objective function as the throughput weighted by the loss visibility of each packet, a proxy for the total perceptual value of successful packets per unit time. We solve the problem of mapping video packets to MIMO subchannels and adapting per-stream rates to maximize the proposed objective. We show that the solution enables jointly reaping gains in terms of improved video quality and lower latency. Optimized packet-stream mapping enables transmission of more relevant packets over more reliable streams while unequal modulation opportunistically increases the transmission rate on the stronger streams to enable low latency delivery of high priority packets. We extend the solution to capture codebook-based limited feedback and MIMO mode adaptation. Results show that the composite quality and throughput gains are significant under full channel state information as well as limited feedback. Tested on H.264-encoded video sequences, for a 4x4 MIMO with 3 spatial streams, the proposed architecture achieves 8 dB power reduction for the same video quality and supports 2.4x higher throughput due to unequal modulation. Furthermore, the gains are achieved at the expense of few bits of cross-layer overhead rather than a complex cross-layer design.


\end{abstract}

\section{Introduction}


The delay-sensitive nature of real-time video transmission motivates the use of unreliable transport protocols, such as user datagram protocol (UDP) for video delivery. This causes the wireless channel impairments, such as losses and delays, to be visible at the APP layer. Consequently, achieving good overall video quality for real-time video requires mitigating channel-induced distortions. Since video quality is the metric of interest from the user perspective, transmission policies should be designed to minimize the impact of losses on video quality. Generally, incorporating video quality-based optimization into lower layer protocols requires a complex, and practically prohibitive, cross-layer design that jointly adapts the video server and the base station. In this paper, we incorporate video quality based optimization into the network without requiring a cross-layer design. Instead, we propose estimating and communicating packet loss visibility and use that measure to optimize video quality. At the cost of few additional bits per packet, video quality-based optimization is enabled by prioritizing video packets at the PHY layer based on perceptual relevance.

The response to video packet losses and distortions is inherently unequal due to the features of state-of-the-art video codecs (e.g. \cite{wiegand2003overview,schwarz2007overview}) such as inter-frame coding, motion compensation, and error concealment. For example, inter-frame coding introduces packet dependencies in the temporal domain, thus causing different error propagation patterns, and increasing the loss visibility variability. Furthermore, the non-uniform motion across different spatial locations causes loss visibility to be unequal across slices and dependent on the error concealment method. Video packet loss visibility captures this unequal response by training a statistical model that maps a set of features per packet to a measure of visibility of that packet loss. More formally, video packet loss visibility is defined as the probability that the artifact due to the loss of a given packet is visible to the average user. The objective of loss visibility modeling and estimation (e.g. \cite{LinPrioritization,kanumuri2006modeling}) is to find the model that best correlates the loss visibility estimate with the results reported by viewers through subjective tests, thus naturally capturing the user perception. Quantizing the loss visibility side information and embedding it into the packet headers enables an inexpensive and effective tool for perceptual quality optimization.


Advanced PHY layer designs, such as multiple-input multiple-output (MIMO) processing, have become an integral part of state-of-the-art wireless standards such as 3GPP Long Term Evolution (LTE) and IEEE 802.11n, which deliver the bulk of stored and real-time video traffic. In this paper, we leverage the spatial degrees of freedom of the MIMO channel to map video packets to MIMO subchannels based on channel quality and packet loss visibility. In short, the proposed technique makes use of the unequal gains of MIMO substreams to provide unequal protection of video packets resulting in a video quality gain. Jointly, unequal modulation is leveraged on the better streams, resulting in a throughput gain and timely delivery of perceptually relevant packets. Consequently, packet prioritization is achieved both in terms of reliability and rate. The major contributions in this paper are summarized as follows.




\subsection{Paper Contributions}



\subsubsection{Low overhead video-aware PHY optimization}

We propose a new low overhead architecture for real-time video transmission to mitigate channel-induced video distortions. Our proposed architecture uses quantized loss visibility scores embedded in the packet header at the expense of only few extra bits per packet while avoiding a complex cross-layer design. We argue that the loss visibility scores of buffered video packets is not sufficient to fully capture the loss visibility variability since real-time video only supports small buffers. Thus, we also estimate the loss visibility distribution inexpensively to capture this variability and provide a notion of relative packet importance that is used in optimizing transmission decisions.

\subsubsection{Packet prioritization for high quality and low latency}

At the PHY layer, we propose to use the loss visibility values to classify video packets into different priority classes. To optimize the loss visibility-based transmission policy for high video quality and low latency, we define an optimization metric that generalizes the conventional notion of throughput by weighting each packet in the optimization objective by its loss visibility. Since loss visibility reflects the visual perception of a corresponding packet loss, our optimization metric is a proxy for \emph{the total perceptual value of packets successfully delivered per unit time}. Given the proposed objective function that enables joint optimization of video quality and latency, we derive optimized PHY layer packet prioritization schemes. We emphasize that the proposed metric is used for optimization rather than evaluation of the algorithm. For assessment of video quality gains, we use objective video quality metrics.

\subsubsection{Loss visibility optimized MIMO precoding}

For a MIMO system, each class of packets is transmitted through a different spatial stream corresponding to a decomposed subchannel of the MIMO channel. We derive the optimal packet-stream mapping that maximizes the loss visibility weighted throughput objective. The solution can be summarized as follows: (1) The MIMO channel is decomposed into parallel streams, (2) the per-stream transmission rate, i.e. modulation order, is chosen to maximize the corresponding throughput per stream, (3) the spatial streams are ordered by their probability of packet error, a function of both the per-stream SNRs and (potentially unequal) modulation orders, (4) the packets are classified according to a thresholding policy whereby higher priority packets are mapped to high order streams as defined by the ordering in (3). The optimal thresholding policy is such that the load is balanced across streams based on the fraction of packet per priority class, the modulation order per stream, and the retransmission overhead. We show that the solution enables jointly reaping gains in terms of improved video quality and lower latency: A packet prioritization gain results from transmission of more relevant packets over more reliable streams and an unequal modulation gain results from opportunistically increasing the transmission rate on the stronger streams to enable low latency delivery of high priority packets.

\subsubsection{Mode adaptation and limited feedback}

We further enhance our algorithm by adapting the MIMO mode corresponding to the number of spatial streams in a manner that jointly captures video quality and throughput maximization. If the loss visibility distribution characterizes a source with high variability, a higher mode is preferable to provide prioritized delivery by adding more packet classes under good channel conditions. Conversely, if the variability in packet importance is low, then the contribution of packet prioritization is minimal and reliable delivery with a smaller number of spatial streams may be preferred. Thus, our proposed approach adapts mode selection according to both the video source and channel conditions. We also extende our Algorithm to codebook-based limited feedback systems where the channel state information is quantized at the receiver and fed back to the transmitter.

\subsection{Related Work}


We review related work on loss visibility estimation and modeling \cite{LinPrioritization,kanumuri2006modeling,kanumuri2006predicting}, loss visibility based optimization \cite{toni2011unequal}, and adaptive MIMO transmission for video content \cite{xu2010mimo,song_videoMIMO,hormis2009adaptive,oyman2010distortion,khalek2012video,khalek2012MM}. In \cite{LinPrioritization}, a generalized linear model is proposed for video packet loss visibility modeling considering factors within a packet and its temporal and spatial vicinity to capture the temporal and spatial distortions. The set of features used to estimate loss visibility is versatile by being applicable over a range of encoding standards, GoP structures, and error concealment methods. Some features such as motion magnitude, motion variance, distance from scene cut, and camera motion capture the video source properties. Other features such as initial structural similarity index (SSIM), maximum per-macroblock (MB) mean square error (MSE), and spatial extent capture the distortions caused by the loss in spatial domain. Temporal error propagation is also captured through features related to the number of frames affected by the loss, distance to reference frame, error concealment method, and other scene loss concealment. The generalized linear model using these features is fit based on subjective tests. Other related loss visibility modeling approaches can be found in \cite{kanumuri2006modeling} and \cite{kanumuri2006predicting}. Besides generalized linear models, \cite{kanumuri2006modeling} proposes a classification-based approach using a statistical tool called classification and regression trees (CART) to classify each packet loss as visible or invisible. The loss visibility model developed in \cite{kanumuri2006modeling} is applied in \cite{toni2011unequal} for selecting unequal coding rates for different slices and for resource allocation in an OFDM system.

In this paper, we propose a generic framework that allows using loss visibility models to optimize transmission policies at the PHY and MAC protocol layers. Specifically, we apply the generalized linear modeling approach in \cite{LinPrioritization} for loss visibility estimation of H.264-encoded sequences due to its versatility and high classification accuracy. We further argue that the loss visibility distribution provides a notion of relative packet importance for real-time video where only a small number of packets are buffered, and thus, we propose to inexpensively estimate and update the distribution using non-parametric learning, and subsequently use it in loss visibility based adaptation.

For scalable video sequences, the loss visibility varies significantly across temporal, spatial, and quality layers. Estimating the average loss visibility of packets from each scalable video layer is addressed in \cite{Khalek_JSAC,Khalek_MLSP,Khalek_Globecom}. Online learning is used to specify the maximum fraction of packet losses from each layer to meet a target video quality. The online algorithm uses local linear regression to estimate the video quality loss due to packet losses from a specific video layer. Based on the ACK history information, the local linear regression fit is updated and the unequal protection levels are estimated continuously over time. Adjusting the learning window provides a tradeoff between factual estimation of loss visibility and finer adaptation to the changing video temporal characteristics.



While loss visibility-based adaptation approaches are not heavily investigated in the literature, other adaptive video transmission techniques such as joint source-channel coding (JSCC) \cite{Girod_JSCC,Kondi_JSCC,zhang2007joint,Khalek_Globecom}, unequal error protection (UEP) \cite{Kim_UEP,Gallant_UEP,Khalek_JSAC}, and prioritized scheduling \cite{Schaar_scheduling}, and distortion-aware resource allocation \cite{Zhang_adaptation,Luo_qualityOptimLTE} have been proposed to increase video quality and error resilience. Previous work, however, does not present a generic framework for incorporating loss visibility-based decisions into wireless networks. To the best of our knowledge, this is the first comprehensive work that defines a generic cross-layer design for using loss visibility in wireless networks, develops MIMO transmission strategies for prioritized delay-sensitive video delivery, and derives corresponding closed-form gain expressions.


Adaptive MIMO transmission for video content has been investigated in \cite{xu2010mimo,song_videoMIMO,hormis2009adaptive,oyman2010distortion}. In \cite{xu2010mimo}, a cross-layer framework for MIMO video broadcast is proposed by allocating scalable video layers to the end-users jointly with precoder computation to ensure that delay and buffer constraints are met. In \cite{song_videoMIMO}, a layered video transmission scheme over MIMO is proposed. It periodically switches each bit stream among multiple antennas to match the ordering of subchannel SNRs, thus providing prioritized delivery. In \cite{hormis2009adaptive}, a method is proposed to adaptively control the diversity and multiplexing gain of a MIMO system to minimize the cumulative video distortion and satisfy delay constraints. Finally, in \cite{oyman2010distortion}, distortion-aware MIMO link adaptation techniques are proposed for MCS and MIMO mode selection. Since \cite{xu2010mimo,song_videoMIMO} are only applicable to scalable video coded bitstreams, the application scope of the proposed techniques is limited as the majority of current video content is non-scalable. Furthermore, \cite{hormis2009adaptive,oyman2010distortion} relies on rate-distortion information which is typically not available for real-time encoded or transcoded video.

\vspace{-0.3cm}

\subsection{Paper Organization and Notation}

The rest of the paper is organized as follows. We present the MIMO system model and the loss visibility-based model in Section \ref{sys_model}. In Section \ref{PT-framework}, we present the background and define the framework for perceptual optimization using loss visibility. In Section \ref{thresholding}, we derive the optimal packet-stream mapping and present the loss visibility optimized MIMO transmission algorithm. In Section \ref{gains}, we derive the corresponding packet prioritization and unequal modulation gains. We present results and analysis using encoded video sequences in Section \ref{results} to quantify the achievable gains. Finally, concluding remarks are provided in Section \ref{conclusion}. Throughout this paper, the following notation is used: $\mathcal{A}$ is a set, $\mathbf{A}$ is a matrix; $\mathbf{a}$ is a vector; and $a$ is a scalar.  The probability density function (PDF) and the cumulative distribution function (CDF) of random variable $A$ are denoted $f_A(.)$ and $F_A(.)$ respectively. Its expectation is denoted by $\mathbb{E}_A\left[.\right]$. We use random variables to characterize the channel variation, determined by the channel matrix, as well as the source variation, determined by the loss visibility values. Other notation is defined when needed.

\section{System Model}\label{sys_model}

This section introduces the proposed MIMO system model that enables loss visibility-based packet prioritization as well as the model for the APP, MAC, and PHY layers.

\subsection{Prioritized MIMO Transmission}\label{sec:prioritized-MIMO}

Consider $P$ packets buffered for transmission where packet $p$ is represented as $\mathbf{s}_p = [s_p[1],\ldots, $ $s_p[b(\mathbf{s}_p)]]$ where $b(\mathbf{s}_p)$ is the number of QAM symbols. The vector of symbols corresponding to all buffered packets is denoted $\mathbf{s} = [\mathbf{s}_1,\ldots,\mathbf{s}_P]^T$.

Consider a  narrowband MIMO wireless system with $N_\mathrm{t}$ transmit antennas and $N_\mathrm{r}$ receive antennas. The system uses $S$ spatial streams where $S\le \min(N_\mathrm{t},N_\mathrm{r})$ and each stream corresponds to a stream of constellation symbols. Our general framework enables the size of the constellation to vary per substream, as well as the number of substreams, known as mode adaptation. Thus, we have $1 \le S \le \min(N_\textrm{t},N_\textrm{r})$. Linear precoding enables mapping a symbol vector from each spatial stream to an $N_\mathrm{t}$-dimensional spatial signal using an $N_\mathrm{t} \times S$ linear precoding matrix $\mathbf{F}_S$. The spatial signal encounters a channel matrix $\mathbf{H}$ and an additive noise vector $\mathbf{n}$ with elements each distributed according to $\mathcal{CN}$(0,$N_0$). The corresponding input-output relationship is

\begin{equation}
\mathbf{y}[i] = \sqrt{\frac{E_\mathrm{s}}{N_\mathrm{t}}}\mathbf{HF}_S \mathbf{T}[i]\mathbf{s} + \mathbf{n}[i]
\end{equation}

\noindent where $\mathbf{y}[i]$ is the received signal and $\mathbf{T}[i]$ is an interleaver matrix that determines the mapping between symbols and spatial streams in the $i^{\textrm{th}}$ channel use and is proposed to enable loss visibility-based prioritized transmission. Note that $\mathbf{T}[i]$ has dimensions $N_t \times \sum_p{b(\mathbf{s}_p)}$. Conventionally, in the absence of loss visibility information, the symbols are transmitted sequentially. Thus, the interleaver for the $i^\mathrm{th}$ channel use can be represented mathematically as

\begin{equation}
\mathbf{T}[i] = \left[\mathbf{0}_{N_t,(i-1)N_t} \mid \mathbf{I}_{N_t} \mid \mathbf{0}_{N_t,\sum_p{b(\mathbf{s}_p) - i N_t}}\right]
\end{equation}

\noindent where $\mathbf{0}_{m,n}$ is an all zeros $m \times n$ matrix and $\mathbf{I}_{m}$ is an $m\times m$ identity matrix. In this paper, we propose designing an interleaver matrix that provides packet prioritization based on loss visibility. Consider a classification policy whereby a set of packets $\mathcal{V}_m$ is classified into priority level $m$ corresponding to packets transmitted through spatial stream $m$. The following interleaver design ensures that packets $p\in\mathcal{V}_m$ are transmitted through stream $m$

\begin{equation}\label{eqn:inter-prioritized}
\mathbf{T}[1]_{m,n} = \left\{ \begin{array}{ll} 1 \textrm{~if~} n = 1 + \sum_{j=1}^{m-1}{\sum_{p\in \mathcal{V}_m}{b(\mathbf{s}_p)}}    \\
0 \textrm{~otherwise~}.\end{array}\right.; \mathbf{T}[i+1] = \left[\mathbf{0}_{N_t,1} \mid \mathbf{T}[i]_{1:N_t,1:\sum_p{b(\mathbf{s}_p)}-1}\right].
\end{equation}

\begin{figure}[t!]
    \centering\hspace{1.2cm}
    \includegraphics[width=\textwidth]{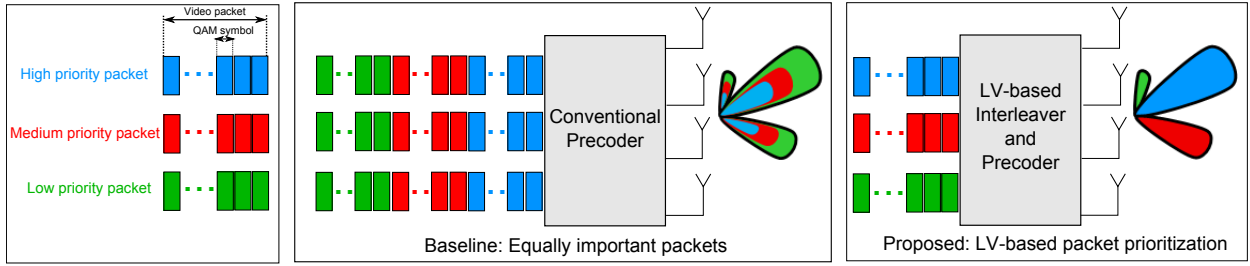}\vspace{-0.3cm}
    \caption{\label{fig:precoder-design} Illustration of the proposed precoder and interleaver design for packet prioritization over MIMO channels.\vspace{-0.3cm}}
\end{figure}

\noindent For practical signal processing purposes, the interleaver matrix in \eqref{eqn:inter-prioritized} is updated inexpensively by ``sliding'' the interleaver from the previous channel use. The resulting mapping is illustrated in Figure \ref{fig:precoder-design} and the physical interpretation of the process is that high priority packets are sent over the more reliable MIMO subchannels.

Given the simple interleaving procedure in \eqref{eqn:inter-prioritized} that enables packet prioritization, the main question we address in subsequent sections is how to determine the classification policy, i.e., given a set of $P$ packets with loss visibility values $\{v(\mathbf{s}_i)\}_{i=1}^{P}$, how to determine the priority sets $\mathcal{V}_i$ to maximize a video quality-based utility function.


\subsection{Precoder Design}

The matrix $\mathbf{HF}_S$ can be thought of as an effective channel. The receiver decodes $\mathbf{y}$ using this effective channel and a zero forcing receiver. We assume a block-fading model whereby the channel realization $\mathbf{H}$ is fixed over a set of packets $\mathcal{P}$ and then independently takes a new realization. All the transmission decisions are adapted every channel coherence time which could be as small as one packet duration, i.e., $P\ge1$, thus being applicable over a range of mobility scenarios. For a zero forcing receiver, it is shown in \cite{heath2001antenna} that the SNR on the $i^\mathrm{th}$ stream is




\vspace{-0.3cm}

\begin{eqnarray}\label{eq:SNR}
\gamma_i (\mathbf{H}) &=& \frac{E_\mathrm{s}}{N_0}\frac{1}{[\mathbf{F}_S^*\mathbf{H}^*\mathbf{H}\mathbf{F}_S]_{i,i}^{-1}}.
\end{eqnarray}

\noindent We consider both cases of perfect and imperfect transmitter channel state information (CSIT). In both scenarios, we assume that the feedback delay is negligible and the transmitter and receiver are fully synchronized. With perfect CSIT, the MIMO channel can be converted to parallel, noninterfering single-input single-output (SISO) channels through a singular value decomposition (SVD) of the channel matrix \cite{telatar1999capacity}. We consider unitary precoding whereby the columns of  $\mathbf{F}_S$ are restricted to be orthogonal. While this could be further generalized to a non-unitary power constraint, we note that using the unitary constraint along with multimode
precoding results in performance near the capacity achieved by waterfilling.  \cite{love2005multimode}. Thus, we create $\mathbf{F}_S$ from a normalized version of the right singular vectors of $\mathbf{H}$ as follows

\vspace{-0.3cm}

\begin{equation}
\mathbf{F}_S = \frac{1}{\sqrt{S}}[\mathbf{V}]_{:,1:S}
\end{equation}

\noindent where $\mathbf{H} = \mathbf{U} \Sigma \mathbf{V}^*$ is the singular value decomposition of $\mathbf{H}$. Under the precoding structure in (3), the SNR for the $i^\mathrm{th}$ stream simplifies to

\vspace{-0.3cm}

\begin{eqnarray}
\gamma_i (\mathbf{H}) &=&\frac{E_\mathrm{s}}{N_0}\frac{\sigma_i^2}{S}
\end{eqnarray}

\noindent where $\sigma_i$ is the $i^\mathrm{th}$ singular value of $\mathbf{H}$. For quantized CSIT, the receiver chooses a precoding matrix $\mathbf{F}_S$ from a codebook $\mathcal{F}_S$ consisting of a finite set of precoding matrices. There are $\log_2(|\mathcal{F}_S|) = B_S$ bits of feedback used to convey the index of the chosen precoding matrix back to the transmitter if $S$ spatial streams is used. For simulations, the codebook $\mathcal{F}_S$ is designed using Grassmannian subspace packing with the chordal subspace distance measure as described in \cite{love2005limited}. The criterion for selecting the precoder at the receiver is to maximize the minimum singular value, that is, $\mathbf{F}_S = \mathrm{argmax}_{\mathbf{F}\in\mathcal{F}}~{\lambda_{\min}(\mathbf{H} \mathbf{F})}$.



\begin{figure}[t!]
    \centering\hspace{1.2cm}
    \includegraphics[width=\textwidth]{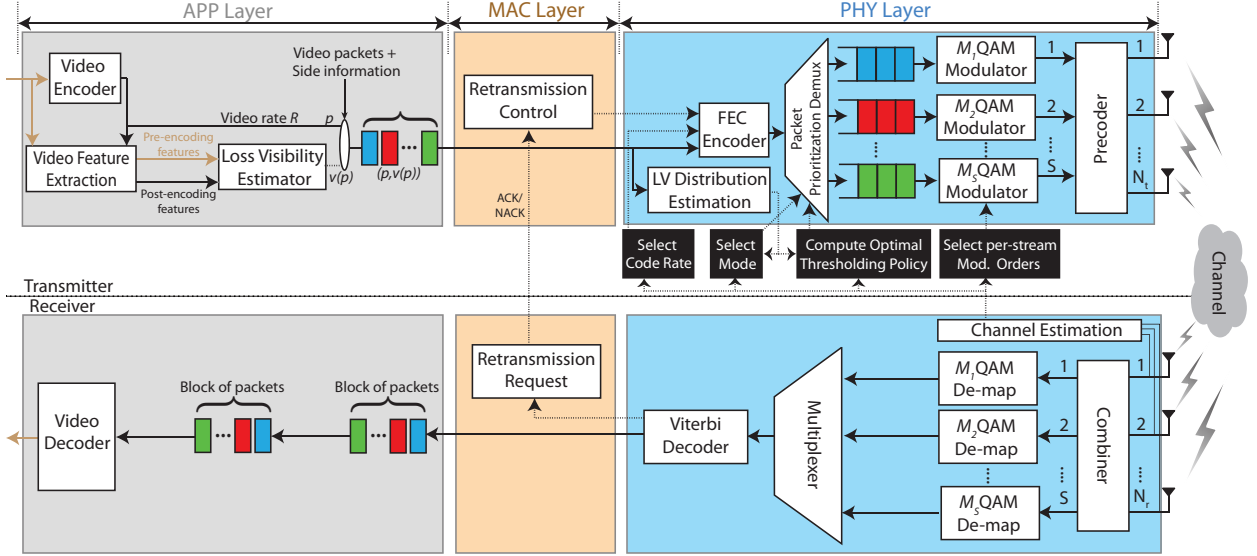}\vspace{-0.3cm}
    \caption{\label{fig:block-diagram} System block diagram for loss visibility based prioritized MIMO transmission.\vspace{-0.3cm}}
\end{figure}



\vspace{-0.3cm}

\subsection{Modulation, Coding, and Retransmission}

We apply unequal modulation per stream. The data through stream $i$ are modulated with a QAM constellation of size $M_i\in \mathcal{M}$ resulting in a data rate $R_i = B \mathrm{log}_2{M_i}$. Each constellation is normalized such that the average symbol energy is unity. For a given channel realization, the vector of modulation schemes is denoted $\mathbf{M} = \{M_i\}_{i=1}^{i=S}$. The set of channel coding rates is $\mathcal{C}$ and the data through all streams are coded with coding rate $C\in \mathcal{C}$.



The probability of packet error through stream $i$ conditioning on the modulation scheme $M_i$, the coding rate $C$, and the $i^\mathrm{th}$ post-processing SNR $\gamma_i (\mathbf{H})$ is denoted $\alpha_i = \mathrm{PER}(M_i,C,\gamma_i (\mathbf{H}))$. While we use the notation $\alpha_i$ for brevity, the dependence on the modulation order, coding rate, and SNR is implied. The uncoded M-QAM error probability expressions $\mathrm{PER}_{\mathrm{uncoded}}(M,\gamma)$ are provided in the literature (e.g. \cite{goldsmith2005wireless}). Given a set of channel codes $\mathcal{C}$, we estimate the coding gain of each particular code as follows. The PER waterfall curve for each MCS $\mathrm{PER}_{\mathrm{uncoded}}(M_i,C,\gamma_i (\mathbf{H}))$ is estimated through Monte-Carlo simulations. Then, the estimated coding gain is the value $g(C)$ that provides the best fit with the translated uncoded expressions, i.e. $g(C) = \mathrm{argmin}\parallel\mathrm{PER}(M,C,\boldsymbol{\gamma})) - \mathrm{PER}_{\mathrm{uncoded}}(M,\boldsymbol{\gamma}+g(C))\parallel$ where $\boldsymbol{\gamma}$ is a representative vector of SNR values. It follows that the coded PER expression can be approximated as $\mathrm{PER}(M_i,C,\gamma_i (\mathbf{H})) \approx \mathrm{PER}_{\mathrm{uncoded}}(M_i,\gamma_i (\mathbf{H})+g(C))$ for each coding rate.



Retransmission with a finite retransmission limit is applied in the system to enable high reliability. Given a retransmission limit of $L$ retransmissions, determined by the MAC protocol, the number of retransmissions follows a truncated geometric distribution assuming the channel is fixed during retransmission. Thus, the mean number of transmissions through stream $i$ is

\begin{eqnarray}
r_i &=& \sum_{k=1}^{L+1}{k(1-\alpha_i)\alpha_i^{k-1}} +(L + 1)\alpha_i^{L + 1} = \frac{1-\alpha_i^{L + 1}}{1-\alpha_i}
\end{eqnarray}

\noindent since $(1-\alpha_i)\alpha_i^{k-1}$ is the probability of success in $k$ transmissions and $\alpha_i^{L + 1}$ is the post-retransmission failure probability. We define the post-retransmission probability of successful packet delivery through stream $i$ as

\vspace{-0.7cm}

\begin{eqnarray}\label{eqn:success-prob}
p^\mathrm{success}_i &=& 1-\alpha_i^{L + 1}.
\end{eqnarray}


\noindent The complete system block diagram including APP layer loss visibility estimation, MAC layer retransmissions and PHY layer packet prioritization is shown in Figure \ref{fig:block-diagram}.

\section{Loss Visibility Estimation and Loss Visibility-based Optimization}\label{PT-framework}

In this section, we first present background on loss visibility estimation and present a framework for using loss visibility side information to characterize the video content. We further propose an optimization metric that uses loss visibility to jointly maximize video quality and network throughput.

\subsection{Background: Loss Visibility Estimation}\label{LV-estimation}

The objective of loss visibility estimation is to associate a packet $p$ with a value $v(\mathbf{s}_p)$ ranging from 0 to 1 and indicating the \emph{loss visibility} of the packet. A value $v(\mathbf{s}_p)=0$ indicates that losing packet $p$ does not have a visible impact on the end video quality whereas a value $v(\mathbf{s}_p)=1$ indicates that the loss of packet $p$ will be visible with probability 1. A PHY packet is composed of one or more slices. If the PHY packet is composed of multiple slices, the packet loss visibility is the mean of the individual slice visibility.

To estimate the loss visibility of APP layer slices, we use the generalized linear model (GLM) approach proposed in \cite{LinPrioritization}. We extract video features both from the raw video reference as well as the encoded bitstream. We note that, for real-time video transmission, the raw video is available at the server since encoding is done in real-time. A video frame is divided into a set of slices, each corresponding to horizontal group of MBs. We apply forward motion estimation to each MB to estimate the motion magnitude for each MB and compute the slice motion magnitude as the average per-MB motion magnitude. The residual energy for each MB is computed from the corresponding motion-compensated residual signal. By thresholding the average motion in the entire video frame, we detect if the scene consists of a still background or if there is camera motion. In addition to these features, we extract features from the encoded bitstream. Specifically, based on the frame type and the inter-frame prediction settings, we flag each packet as affecting one or multiple frames. To capture spatial-domain distortions, we further compute the initial SSIM feature corresponding to the SSIM in the frame affected by the loss, and max initial mean square error (IMSE) representing maximum per-MB MSE in the same frame. For videos sequences with multiple scenes, we detect scene cuts and use that to flag packets concealed using a reference  corresponding a previous scene for which losses are more visible. We also flag packets before scene cuts for which losses will be barely visible. Scene cuts are detected simply by comparing the residual energy between each two consecutive frames to a preset threshold. While other features are defined in \cite{LinPrioritization}, subjective tests show that only the ones mentioned above have high (positive or negative) correlation with loss visibility as reported by viewers. Using all these features, we use the following logistic regression model for loss visibility estimation

\begin{equation}
\log\left(\frac{v(\mathbf{s}_p)}{1-v(\mathbf{s}_p)}\right) = \beta_0 + \sum_{i=1}^{F}{\beta_i x_{pi}}
\end{equation}

\noindent where $\boldsymbol{\beta} = \{\beta_0,\beta_1,\ldots,\beta_F\}$ are the intercept and the coefficients associated with the different features. We use the coefficients as reported in Table IV in \cite{LinPrioritization}. We assume the loss visibility $v(\mathbf{s}_p)$ of packet $p$ is communicated to the physical layer through the packet header and deep packet inspection can be performed at the network edge to extract the loss visibility.

Our system model allows for unequal packet sizes and the packet value $v(\mathbf{s}_p)$ and the packet size $b(\mathbf{s}_p)$ may in general be correlated, as is the case in practice. We assume, however, that if $v(\mathbf{s}_1)>v(\mathbf{s}_2)$, then $b(\mathbf{s}_1)>b(\mathbf{s}_2)$. This is typically the case since low visibility packets (e.g. B frame packets) are predictively encoded, and thus compressed more efficiently.



\subsection{Loss Visibility Distribution Estimation}



Over a sufficiently long timescale, the distribution of the loss visibility values characterizes the video source and the codec. For instance, a GoP structure $IBPBP\cdots$ results in a larger concentration of low visibility packets than $IPPPP\cdots$. Thus, we estimate the loss visibility distribution to be used in packet classification. We propose to estimate the \emph{loss visibility distribution} using kernel density estimation (KDE) \cite{scott1992multivariate}, update it using the values of incoming packets, and use it to derive the optimal packet prioritization policy. With KDE, the density estimate at $0\le x\le 1$, denoted by $\hat{f}_v(x)$, is

\begin{equation}\label{eq:KDE}
\hat{f}_v(x) = \frac{1}{W}\sum_{i=1}^{W}{K_h(x - v(\mathbf{s}_{p-i}))} = \frac{1}{W h}\sum_{i=1}^{W}{K\left(\frac{x - v(\mathbf{s}_{p-i})}{h}\right)}
\end{equation}

\noindent where $W$ is the window corresponding to the number of packets over which the estimate is obtained and $K_h(\cdot)$ is a kernel with smoothing parameter $h > 0$. Adjusting the kernel density estimation window $W$ and smoothing parameter $h$ provides a bias/variance tradeoff between factual estimation of the loss visibility and fine adaptation to changing video characteristics. The distribution is inexpensive to compute and update as it only consists of a linear operations.

The main advantage of using the loss visibility distribution is that for real-time video, where large buffers are not available, the buffered packet values are not fully representative of the loss visibility variability. Thus, the loss visibility distribution is used instead to capture this variability and provide a notion of relative packet importance.

\subsection{Loss Visibility-Weighted Throughput: An Optimization Metric}\label{PT-metric}

To jointly capture the two desirable objectives of high video quality and low latency video delivery, we propose optimizing throughput weighted by per-packet loss visibility. This generalizes the conventional notion of throughput to unequally important packets. Maximizing loss visibility-weighted throughput is equivalent to \emph{maximizing the total perceptual value of packets successfully delivered per unit time}. This enables composite gains in perceived video quality and throughput. The loss visibility-weighted throughput expression is

\begin{eqnarray}\label{eq:PT-general}
{WT} &=& \frac{\sum_{v}{q^\mathrm{success}(v)v }}{t(\mathbf{H},\mathbf{M},C,\{\mathcal{V}_i\}_{i=1}^{S})}
\end{eqnarray}



\noindent where $q^\mathrm{success}(v)$ is the probability that a packet with loss visibility $v$ is successfully delivered (after potential retransmission), and $t(\mathbf{H},\mathbf{M},\{\mathcal{V}_i\}_{i=1}^{S})$ is the time to transmit the packets given the packet-stream mapping $\{\mathcal{V}_i\}_{i=1}^{S}$, the channel matrix $\mathbf{H}$, modulation orders $\mathbf{M}$, and the coding rate $C$. The dependence of the success probability on the packet values is intended to capture general unequal error protection policies. In the proposed packet prioritization policy presented in \S\ref{sec:prioritized-MIMO}, the expression reduces to

\begin{eqnarray}\label{eq:PT-general}
{WT}_{\textrm{prioritized}} &=& \frac{\sum_{i=1}^{S}{p^\mathrm{success}_i(\gamma_i (\mathbf{H}),\mathbf{M},C)~ \sum_{v\in\mathcal{V}_i}{ v  }}}{\max_i t_i(\gamma_i (\mathbf{H}),M_i,C,\mathcal{V}_i)}
\end{eqnarray}


\noindent since $q^\mathrm{success}(v) = p^\mathrm{success}_i(\gamma_i (\mathbf{H}),\mathbf{M},C) = 1-\alpha_i^{L+1}$ is the probability of post-retransmission successful packet delivery defined in \eqref{eqn:success-prob} if $v\in\mathcal{V}_i$. Alternatively, for the baseline where no loss visibility side information is used, the loss visibility-weighted throughput expression is

\begin{eqnarray}\label{eq:PT-baseline}
{WT}_{\textrm{baseline}} &=& \frac{p^\mathrm{success}_\mathrm{baseline}(\mathbf{H},\mathbf{M},C)~ \sum_{v\in\cup_i \mathcal{V}_i}{ v  }}{ t(\mathbf{H},\mathbf{M},C)}
\end{eqnarray}

\noindent where $p^\mathrm{success}_\mathrm{baseline}(\mathbf{H},\mathbf{M},C) = 1-\alpha^{L+1}_\mathrm{baseline}$ for the baseline case whereby each packet is multiplexed over all streams. We note that packet error rate in the baseline case $\alpha_\mathrm{baseline}$ and the prioritized transmission case $\alpha_i$ can be related as follows. Consider a packet of $b$ QAM symbols with a symbol error rate $\mathrm{SER}_i$ through stream $i$, in the prioritized transmission scenario, the packet error rate corresponding to transmission through stream $i$ is $\alpha_i = 1 - (1-\mathrm{SER}_i)^{b}$. Alternatively, without packet prioritization, the packet is transmitted over $b/S$ channel uses through all streams and the corresponding packet error rate is $\alpha_\mathrm{baseline} = 1 - \prod_i{(1-\mathrm{SER}_i)^{b/S}}$. Substituting for $\alpha_i$, we obtain

\begin{equation}
\alpha_\mathrm{baseline} = 1-\prod_{i=1}^{S}{(1-\alpha_i)}^{1/S}.\label{eq:overall-PER}
\end{equation}

\noindent


\begin{table}[t!]\footnotesize\label{tbl:notation}
  \caption{Commonly used notation}\vspace{-0.2cm}\centering
\begin{tabular}{|c|l|}
  \hline
  $N_t$ & Number of transmit antennas \\
  $N_r$ & Number of receive antennas \\
  $S$ & Number of spatial streams \\\hline
  \hline
  $f_v(v)$ & Packet loss visibility distribution\\
  $V_i$ & Cumulative loss visibility values of class $i$ packets (i.e., transmitted through the $i^{\mathrm{th}}$ stream)\\
  $\hat{\mathbf{v}} = \{\hat{v}_i\}_{i=2}^S$ & Vector of loss visibility thresholds where $\hat{v}_i$ is the threshold between stream $i$ and $i-1$ \\
\hline \hline
  $\gamma_i (\mathbf{H})$ & Post-processing SNR on $i^{\mathrm{th}}$ stream\\
  $t_i$ & Mean time to transmit a class $i$ packet\\
$\mathbf{M} = \{M_i\}_{i=1}^{i=S},~M_i\in\mathcal{M}$& Vector of modulation schemes per stream\\
$R_i = B \log_2(M_i)$& Data rate on stream $i$\\
$C\in \mathcal{C}$&Coding rate\\
  $\alpha_i = \mathrm{PER}(M_i,C,\gamma_i (\mathbf{H}))$ & Packet error rate for packets transmitted through stream $i$\\
  $\alpha_\mathrm{baseline} = \mathrm{PER}(M_i,C,\gamma_i (\mathbf{H}))$ & Packet error rate for packets multiplexed through all streams (baseline) \\
$p^\mathrm{success}_i = 1-\alpha_i^{L+1}$ & Post-retransmission probability of successful packet delivery through stream $i$\\
$p^\mathrm{success}_\mathrm{baseline} = 1-\alpha_\mathrm{baseline}^{L+1}$ & Post-retransmission probability of success by multiplexing through all streams (baseline)\\
  $r_i$ & Average number of retransmissions for packets transmitted through stream $i$\\
    \hline
\end{tabular}\vspace{-0.5cm}
\end{table}

\section{Loss Visibility-based Packet Prioritization}\label{thresholding}

In this section, we formulate the prioritized video transmission problem over MIMO channels and we derive the optimal packet prioritization policy that maximizes the loss visibility-weighted throughput.

\subsection{Problem Formulation}

We propose to solve the problem

\vspace{-0.7cm}

\begin{eqnarray}
&\mathrm{max}_{\{\mathcal{V}_i\},\mathbf{M},C,S} & {WT}_{\textrm{prioritized}}(\{\mathcal{V}_i\},\mathbf{M},C,S) \label{eqn:quality-obj}
\\&\mathrm{s.t.} & \cup_{i=1}^{S} \mathcal{V}_i = [0,1] \label{eqn:constr2}
\\&& M_i\in\mathcal{M} ~\forall i=1,\ldots,S;~C\in\mathcal{C}.\label{eqn:constr5}
\end{eqnarray}

\noindent The objective is to select the number of packet classes $S$ and the classification policy determining the mapping of the set of packets $\mathcal{V}_i$ to spatial stream $i$, as well as the modulation orders $\mathbf{M}$ and the coding rate $C$ such that the weighted throughput objective is maximized.

\subsection{Stream Ordering}\label{sec:order}

First, we show that the set $\mathcal{V}_i$ that maximizes the proposed weighted throughput objective has a simple form obtained by ordering the spatial streams by the corresponding probability of error and mapping the packets onto the ordered streams according to a set of thresholds.


\begin{Lemma1}
The optimal packet-stream mapping is such that $\mathcal{V}_{i}$ has the form $\mathcal{V}_{i}=[\hat{v}_i,\hat{v}_{i+1}]$ where $\cup_{i=1}^{S} \mathcal{V}_{i} = [0,1]$. Furthermore, for any two packets $\mathbf{s}_1$ and $\mathbf{s}_2$ s.t. $v(\mathbf{s}_1)<v(\mathbf{s}_2)$, $\mathbf{s}_1\in \mathcal{V}_i$ and $\mathbf{s}_2\in \mathcal{V}_{k}$ where $p^\mathrm{success}_i\le p^\mathrm{success}_{k}$. It follows that the streams should be ordered by the probability of success $p^\mathrm{success}_i \le p^\mathrm{success}_{i+1}$.
\end{Lemma1}

\begin{proof} See Appendix A.\end{proof}

Note that the ordering in Lemma 1 captures the effect of modulation, coding, retransmission, and channel state because $p^\mathrm{success}_i$ is a function of $M_i$, $C$, $r$, and $\boldsymbol{\gamma}(\mathbf{H})$. In fact, the result represents a generalization of SNR ordering to the case of unequal modulation per stream.

The classification policy reduces into a thresholding policy completely determined by the vector of thresholds $\hat{\mathbf{v}} = \{\hat{v}_i\}_{i=1}^{i=S+1}$. Furthermore, the constraint in \eqref{eqn:constr2} can be rewritten as $0\le\hat{v}_i\le\hat{v}_{i+1}\le 1$ where $\hat{v}_{1} = 0$ and $\hat{v}_{S+1} = 1$ by definition. Thus, we have

\begin{eqnarray}\label{eq:PT-general}
{WT}_{\textrm{prioritized}} &=& \frac{\sum_{i=1}^{S}{p^\mathrm{success}_i(\gamma_i (\mathbf{H}),\mathbf{M},C)~ \sum_{p\in\mathcal{V}_i}{ v(\mathbf{s}_p)  }}}{\max_i t_i(\gamma_i (\mathbf{H}),M_i,C,\mathcal{V}_i)}.
\end{eqnarray}

\noindent Now, we expand \eqref{eq:PT-general} by writing $t_i(\gamma_i (\mathbf{H}),M_i,C,\mathcal{V}_i)$ in terms of the respective parameters. The time to transmit a packet through stream $i$ is $ b(\mathbf{s}_p) (1-\alpha_i^{L + 1}) / (C R_i (1-\alpha_i))$ where $b(\mathbf{s}_p)$ is the size of packet $p$. Taking the expectation over class $i$ packets, we obtain



\vspace{-0.3cm}

\begin{eqnarray}
t_i(\gamma_i (\mathbf{H}),M_i,C,\mathcal{V}_i) &=& \mathbb{E}\left[\frac{b(\mathbf{s}_p) (1-\alpha_i^{L + 1})}{ C R_i (1-\alpha_i)}\right](F_v(\hat{v}_{i+1})-F_v(\hat{v}_{i}))\nonumber \\&=& \frac{\mathbb{E}[b(\mathbf{s}_p)] (1-\alpha_i^{L + 1})}{C R_i (1-\alpha_i)}(F_v(\hat{v}_{i+1})-F_v(\hat{v}_{i}))
\end{eqnarray}

\noindent where $\mathbb{E}[b(\mathbf{s}_p)]$  is the mean packet size. Thus, the weighted throughput expression is

%

\begin{eqnarray}
\hspace{-0.5cm}{WT}_{\textrm{prioritized}}(\hat{\mathbf{v}},\mathbf{M},C,S) \hspace{-0.3cm}&=& \hspace{-0.3cm}\frac{\left[\sum_{i=1}^{S}{(1-\alpha_i^{L + 1}) \int_{\hat{v}_{i}}^{\hat{v}_{i+1}}{v f_v(v) \mathrm{d}v}}\right]}{\mathbb{E}[b(\mathbf{s}_p)]\max_{i}\{(F_v(\hat{v}_{i+1})-F_v(\hat{v}_{i}))(1-\alpha_i^{L + 1})/C R_i (1-\alpha_i)\}}\label{eq:T_R_S1}
\\&=& \hspace{-0.3cm}\underbrace{\frac{C R_{\tilde{i}}(1-\alpha_{\tilde{i}})}{\mathbb{E}[b(\mathbf{s}_p)] (1-\alpha_{\tilde{i}}^{L + 1})}}_{\textrm{Throughput component}}\underbrace{\frac{\sum_{i=1}^{S}{(1-\alpha_i^{L + 1}) \int_{\hat{v}_{i}}^{\hat{v}_{i+1}}{v f_v(v) \mathrm{d}v}}}{(F_v(\hat{v}_{\tilde{i}+1})-F_v(\hat{v}_{\tilde{i}}))}}_{\textrm{Video quality component}}\label{eq:T_R_S2}
\end{eqnarray}

\noindent where $\tilde{i} = \mathrm{argmax}_i\{(F_v(\hat{v}_{i+1})-F_v(\hat{v}_{i}))(1-\alpha_i^{L + 1})/C R_i (1-\alpha_i)\}$ denotes the stream with the longest transmission time on average.


\subsection{Optimal Thresholding Policy: A Load Balancing Solution}

In this section, we derive the optimal thresholding policy $\hat{\mathbf{v}}^*$ for any continuous loss visibility distribution given the optimal ordering in \S\ref{sec:order}.

In Lemma 2 and Lemma 3, we derive properties of the gradient $\partial {WT}_{\textrm{prioritized}}/\partial \hat{v}_i$ that will be used to find the thresholds $\hat{v}_i$ that maximize the weighted throughput expression in Theorem 1.

\begin{Lemma2}
If the streams are ordered by the post-retransmission success probability, i.e., $p^\mathrm{success}_i \le p^\mathrm{success}_{i+1}~\forall i = 1,\cdots, N_s - 1$, then the gradient $\partial {WT}_{\textrm{prioritized}}/\partial \hat{v}_i$ satisfies the following properties:

\begin{enumerate}
\item $\partial {WT}_{\textrm{prioritized}}/\partial \hat{v}_{\tilde{i}}\ge0$ where $\tilde{i} = \mathrm{argmax}~t_i$
\item $\partial {WT}_{\textrm{prioritized}}/\partial \hat{v}_i\le0~\forall i\ne \tilde{i}$
\end{enumerate}

\end{Lemma2}

\begin{proof} See Appendix B. \end{proof}

\vspace{0.3cm}

We use Lemma 2 to derive a more general condition on the behavior of the gradient for the case where $\exists \tilde{j} \ne \tilde{i} \mathrm{~s.t.~} \tilde{i} = \tilde{j} = \mathrm{argmax}~t_i$, i.e., more than one stream have the same average transmission time. This extension will be key to proving the result in Theorem 1.

\begin{Lemma3}Define $\mathcal{I} = \{\mathrm{argmax}~t_i\}$. If $\{\hat{v}_{i}; i\in\mathcal{I}$ or $i-1\in\mathcal{I}\}$ are jointly scaled to keep $\mathcal{I}$ fixed, then

\begin{enumerate}
\item $\partial {WT}_{\textrm{prioritized}}/\partial \hat{v}_{i}\ge0$ if $i\in\mathcal{I}$ and $i-1\not\in\mathcal{I}$
\item $\partial {WT}_{\textrm{prioritized}}/\partial \hat{v}_i\le0$ if $i\not\in\mathcal{I}$ and $i-1\in\mathcal{I}$
\end{enumerate}

\end{Lemma3}

\begin{proof} See Appendix C. \end{proof}

\vspace{0.3cm}


Theorem 1 provides the optimal thresholding policy among streams and applies for any continuous loss visibility distribution obtained using kernel density estimation based on \eqref{eq:KDE}.

\begin{Thm1}\textbf{Thresholding Policy:} The optimal loss visibility thresholds $\hat{\mathbf{v}}^* = \{\hat{v}_{i}^*\}_{i=2}^{S}$ satisfy

\begin{equation}
F_v(\hat{v}_{i+1}^*) - F_v(\hat{v}_i^*) = \frac{R_{i}/r_i }{\sum_{j=1}^{S}{ R_j/r_j}}~\forall i = 1,\cdots,S \label{eq:optimal_splitting}
\end{equation}

\noindent where $r_i = (1-\alpha_i^{L + 1})/(1-\alpha_i)$.
\end{Thm1}

\begin{proof} See Appendix D. \end{proof}

%

The solution is such that the post-retransmission throughput is equal among streams. Thus, the thresholds are selected to balance the load among spatial streams in proportion to the achievable throughput on each stream and the corresponding fraction of packets in each of the $S$ classes. Correspondingly, the solution is referred to as the load balancing solution.



Figure \ref{fig:Thresholding} illustrates the result for a specific channel realization and the Foreman video sequence. First, we show the loss visibility distribution obtained using kernel density estimation. Next, the MIMO channel is decomposed to obtain $\boldsymbol{\gamma}(\mathbf{H})$. Based on the SNR per stream, the throughput-maximizing constellation is chosen per stream. Given the loss visibility distribution, the constellation order, and the corresponding packet error rate, the set of thresholds are determined.  The most prominent result in Figure \ref{fig:Thresholding} is that the \emph{high priority packets are sent with higher reliability (lower packet error rate / retransmission overhead) and lower latency (higher order constellation)}. Thus, utilizing the MIMO channel structure in the manner proposed enables both fewer errors and lower latency for the video packets that matter most making it particularly suitable for real-time video.

We further emphasize the cross-layer nature of the solution based on the components of \eqref{eq:optimal_splitting} in the following three aspects:


\begin{figure}[t!]
    \centering\hspace{1.2cm}
    \includegraphics[height=3cm]{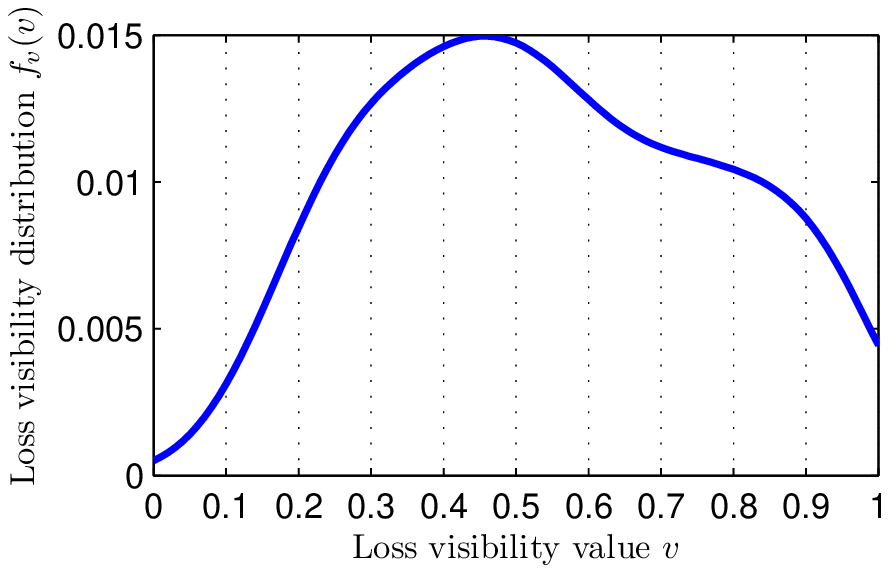}\raisebox{1.3cm}{\scalebox{2}{$\Huge\rightarrow$}}
    \includegraphics[height=3.4cm]{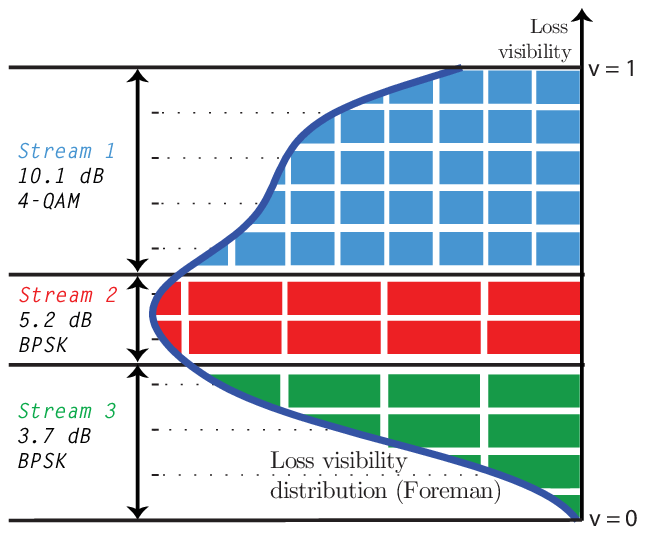}\raisebox{1.3cm}{\scalebox{2}{$\Huge\rightarrow$}}
    \includegraphics[height=2.6cm]{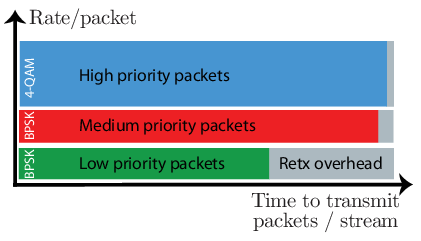}\caption{\label{fig:Thresholding}Graphical illustration of loss visibility optimized transmission policy for $\boldsymbol{\gamma} (\mathbf{H}) = [10.1;5.2;3.7]$ dB and the Forman video sequence; (a) Obtain loss visibility distribution (shown for the Foreman video sequence), (b) Decompose MIMO channel, (c) Determine throughput-maximizing modulation order per stream, (d) Find the optimal thresholding policy. Note that high priority packets achieve both higher rate and reliability. \vspace{-0.7cm}}
\end{figure}

\begin{enumerate}
\item \emph{Non-uniform loss visibility distribution (\textbf{APP}):} The loss visibility thresholds are selected to balance the fraction of packets through each stream based on the loss visibility distribution. In Figure \ref{fig:Thresholding}, this can be seen on the second stream where $\hat{v}_3 - \hat{v}_2$ is made small enough to compensate for the larger concentration of medium priority packets so that the load is balanced among streams.

\item \emph{Unequal modulation per stream (\textbf{PHY}):} If the SNR on spatial stream $i$ allows supporting a higher modulation order $M_i$, the fraction of packets through stream $i$ is increased accordingly. In Figure \ref{fig:Thresholding}, this can be seen on the uppermost stream.

\item \emph{Retransmission overhead (\textbf{MAC}):} If a particularly low SNR on spatial stream $i$ incurs a large retransmission overhead $r_i$, the fraction of packets through stream $i$ is reduced accordingly. In Figure \ref{fig:Thresholding}, this can be seen on the lowermost stream.
\end{enumerate}

Under the load balancing solution in Theorem 1, we have

\begin{equation}
{WT}_{\textrm{prioritized}}(\hat{\mathbf{v}}^*,\mathbf{M},C,S) = \underbrace{\frac{C}{\mathbb{E}[b(\mathbf{s}_p)]}\sum_{i=1}^{S}{\frac{1-\alpha_i}{1-\alpha_i^{L + 1}}R_i }}_{\mathrm{Post-retx~sum~throughput}} \underbrace{\left[\sum_{i=1}^{S}{(1-\alpha_i^{L + 1}) \int_{\hat{v}_{i}}^{\hat{v}_{i+1}}{v f_v(v) \mathrm{d}v}}\right]}_{\mathrm{Loss-penalized~quality~measure}}. \label{eq:PT_optimal_v}
\end{equation}

\noindent We note that for the special case of full retransmission, i.e., $L=\infty~\forall i$, \eqref{eq:PT_optimal_v} reduces to the sum throughput as follows

\begin{equation}
{WT}_{\textrm{prioritized}}(\hat{\mathbf{v}}^*,\mathbf{M},C,S) =   \frac{C}{\mathbb{E}[b(\mathbf{s}_p)]}\sum_{i=1}^{S}{(1-\alpha_i) R_i}. \label{eq:PT_optimal_v_fullretx}
\end{equation}

\noindent In this limiting case, where infinite retransmissions are allowed, all packets are eventually delivered reliably and providing packet prioritization on the basis of video quality becomes obsolete. Thus, the objective function reduces to throughput optimization.

\subsection{MIMO Mode Selection and Link Adaptation}\label{mode-selection}

Next, we discuss the selection of the modulation order per stream, the coding rate, and the MIMO mode to optimize the target objective. Link adaptation enables adapting the modulation and coding to the channel conditions. We optimize the modulation and coding order to maximize the throughput component of \eqref{eq:PT_optimal_v}. Thus, we have

\begin{eqnarray}
\{\mathbf{M}^*,C^*\} &=& \mathrm{argmax}_{M_i\in\mathcal{M},C\in\mathcal{C}}\left\{C\sum_i{ \frac{1-\alpha_i(\gamma_i,M_i,C)}{1-\alpha_i(\gamma_i,M_i,C)^{L+1}}R_i}\right\}\\
&=& \mathrm{argmax}_{C\in\mathcal{C}}\left\{C\sum_i{\mathrm{argmax}_{M_i\in\mathcal{M},C\in\mathcal{C}}\left\{ \frac{1-\alpha_i(\gamma_i,M_i,C)}{1-\alpha_i(\gamma_i,M_i,C)^{L+1}}R_i\right\}}\right\}.
\end{eqnarray}

\noindent Thus, the optimal modulation and coding combination can be found in the following two steps. First, for any given code rate, a corresponding set of modulation orders per stream are selected as follows

\begin{eqnarray}
\tilde{M}_i(C) &=& \mathrm{argmax}_{M_i\in\mathcal{M}}\left\{ \frac{1-\alpha_i(\gamma_i,M_i,C)}{1-\alpha_i(\gamma_i,M_i,C)^{L+1}}R_i \right\}.
\end{eqnarray}

\noindent Next, given $\{\tilde{M}_i(C)\}_{i=1}^{S}$ for every $C\in\mathcal{C}$, we select the optimal code rate and the corresponding optimal modulation order per stream as follows

\begin{eqnarray}
C^* &=& \mathrm{argmax}_{C\in\mathcal{C}}\left\{C\sum_i{\log_2 \tilde{M}_i(C) \frac{1-\alpha_i(\gamma_i,M_i,C)}{1-\alpha_i(\gamma_i,M_i,C)^{L+1}}}\right\},\nonumber\\
M_i^* &=& \tilde{M}_i(C^*)~\textrm{and}~R_i^* = B\log_2(M_i^*) .\nonumber
\end{eqnarray}

\noindent Substituting $\{\mathbf{M}^*,C^*\}$ into \eqref{eq:PT_optimal_v}, we obtain

\begin{equation}
{WT}_{\textrm{prioritized}}(\hat{\mathbf{v}}^*,\mathbf{M}^*,C^*,S) =   \frac{C^*}{\mathbb{E}[b(\mathbf{s}_p)]}{\sum_{i=1}^{S}{\frac{1-\alpha_i}{1-\alpha_i^{L + 1}}R_i^* }\left[\sum_{i=1}^{S}{(1-\alpha_i^{L + 1}) \int_{\hat{v}_{i}}^{\hat{v}_{i+1}}{v f_v(v) \mathrm{d}v}}\right]}.\label{eq:PT_optimal_v_MCS}
\end{equation}

\noindent Practical MIMO link adaptation should include a mechanism for switching the mode, i.e., the number of spatial streams based on channel state matrix $\mathbf{H}$ to optimize system performance and provide a suitable diversity-multiplexing tradeoff. This allows a continuum of operating points that provide different data rate and reliability combinations from single stream beamforming to full spatial multiplexing. In this work, the MIMO mode selection criterion is intended to capture video quality as well as throughput. On one hand, if the loss visibility distribution experiences higher variability, it may be preferable to use more streams to provide prioritized delivery by adding more packets classes if the channel quality is good. On the other hand, if the variability in packet importance is low, then the contribution of packet prioritization is minimal and reliable delivery with a smaller number of spatial streams may be preferred. Thus, mode selection can adapt according to the video source in a content-aware manner. Consequently, the mode selection criterion is to maximize the weighted throughput expression:

\begin{equation}\label{eq:mode-selection}
S^*  = \left\{\mathrm{argmax}~ {WT}_{\textrm{prioritized}}(\hat{\mathbf{v}}^*,\mathbf{M},C,S)\mathrm{~s.t.~} C^*\sum_{i=1}^{S}{\frac{1-\alpha_i}{1-\alpha_i^{L + 1}}R_i^* } >R\right\}
\end{equation}

\noindent where $R$ is the video source rate. The constraint $ C^*\sum_{i=1}^{S}{{(1-\alpha_i)R_i^* }/(1-\alpha_i^{L + 1})} >R$ ensures the throughput with the selected mode at least matches the rate of the video to ensure that the wireless link can serve the requirements the video source.



\subsection{Loss Visibility Optimized Video Transmission Algorithm}\label{Algorithm}

In this section, we describe the proposed algorithm for loss visibility-optimized video transmission over MIMO systems which involves selecting the optimal thresholding policy and the MCS per stream given the post-processing SNRs corresponding to the MIMO channel decomposition.

The algorithmic description is provided in Algorithm 1. Given a certain number of spatial streams $S$, the algorithm computes the corresponding precoder $\mathbf{F}_S$ to maximize the minimum singular value and the corresponding post processing SNRs per stream. It then selects the modulation orders to maximize the per-stream throughput and the coding rate to maximize the overall throughput.

The algorithm orders the streams according to the post-retransmission success probability. Given the modulation orders per stream and the loss visibility distribution, the optimal thresholding policy is computed according to Theorem 1. This determines the values of the thresholds for transmission through each stream. After the process is repeated for each mode, the mode that maximizes the objective function and supports the video source rate is chosen according to \eqref{eq:mode-selection}. The block of packets corresponding to a channel coherence time are transmitted according to the selected MCSs, thresholding policy, and MIMO mode. Given the values of the incoming packets, the algorithm updates the estimated loss visibility distribution using kernel density estimation at each channel coherence time.

\begin{boxedalgorithmic}{1}{Loss Visibility Optimized Video Transmission over MIMO.}

\STATE Given channel state $\mathbf{H}$

\FOR{$i = 1 \to S$}

\STATE \textbf{\emph{Step 1. Precoder Computation}}

\STATE Compute precoder $\mathbf{F}_S$ and post-processing SNRs $\boldsymbol{\gamma}(\mathbf{H}) = \{\gamma_i(\mathbf{H})\}_{i=1}^{S}$

\STATE \textbf{\emph{Step 2. MCS Selection}}

\FOR{$C \in \mathcal{C}$}

\STATE $\tilde{M}_i(C) = \mathrm{argmax}_{M_i\in\mathcal{M}}\left\{ \frac{1-\alpha_i}{1-\alpha_i^{L+1}}R_i \right\}$

\ENDFOR

\STATE $C^* = \mathrm{argmax}_{C\in\mathcal{C}}\left\{C\sum_i{\log_2 \tilde{M}_i(C) \frac{1-\alpha_i}{1-\alpha_i^{L+1}}}\right\}$

\STATE $M_i^* = \tilde{M}_i(C^*)$

\STATE Order streams according to post-retransmission success probability, i.e., $p^\mathrm{success}_i \le p^\mathrm{success}_{i+1}~\forall i=1,\cdots,S-1$.

\STATE \textbf{\emph{Step 3. Loss Visibility Distribution Update}}

\STATE Use kernel density estimation to update the loss visibility distribution $\hat{f}_v(x) = \frac{1}{W h}\sum_{i=1}^{W}{K\left(\frac{x - v(\mathbf{s}_{p-i})}{h}\right)}$

\STATE \textbf{\emph{Step 4. Thresholding Policy Selection}}

\STATE Compute $\hat{\mathbf{v}}^* = \{\hat{v}_{i}^*\}_{i=2}^{S}$ to satisfy $F_v(\hat{v}_{i+1}^*) - F_v(\hat{v}_i^*) = \frac{R_{i}/r_i }{\sum_{j=1}^{S}{ R_j/r_j}}~\forall i = 1,\cdots,S $

\ENDFOR

\STATE \textbf{\emph{Step 5. Mode Selection}}

Select the optimal mode $S^*$ according to \eqref{eq:mode-selection}.

\end{boxedalgorithmic}

\section{Video Quality and Throughput Gains}\label{gains}

To quantify the gains from using the loss visibility side information as proposed in Algorithm 1, we compare with conventional MIMO transmission whereby no side information is used for packet prioritization. Instead, the symbols corresponding to each packet are multiplexed on all spatial streams.



\subsection{Gain Analysis}

In the absence of packet prioritization, each packet is multiplexed over all streams, thus, the packet error rate expression should be modified to capture the new packet error rate. As shown in \eqref{eq:overall-PER}, the corresponding PER relates to the packet prioritization case as follows $\alpha_\mathrm{baseline} = 1-\prod_{i=1}^{S}{(1-\alpha_i)}^{1/S}$. Further, the probability of success for the baseline case is expressed as $p^\mathrm{success}_\mathrm{baseline}(\mathbf{H},\mathbf{M},C) = (1-\alpha^{L+1}_\mathrm{baseline})$. Thus, for a representative set of $P$ packets, the cumulative value of packets received successfully is $P(1-\alpha_\mathrm{baseline}^{L+1}) \mathbb{E}[v(\mathbf{s}_p)]$ where $\mathbb{E}[v(\mathbf{s}_p)] = \int_{0}^{1}{v f_v(v) \mathrm{d}v}$ is the average packet loss visibility. Furthermore, the transmission time is $\max_{i}\{\mathbb{E}[b(\mathbf{s}_p)](1-\alpha_\mathrm{baseline}^{L+1})/C R_i(1-\alpha_\mathrm{baseline})\}P/S = \mathbb{E}[b(\mathbf{s}_p)](1-\alpha_\mathrm{baseline}^{L+1})P/S(1-\alpha_\mathrm{baseline})C \min_{i}\{R_i\}$. Thus, the weighted throughput objective for the baseline follows from \eqref{eq:PT-baseline} as follows

\begin{eqnarray}
{WT}_{\textrm{baseline}} &=& \frac{P(1-\alpha_\mathrm{baseline}^{L+1}) \mathbb{E}[v(\mathbf{s}_p)]}{\mathbb{E}[b(\mathbf{s}_p)](1-\alpha_\mathrm{baseline}^{L+1})P/S(1-\alpha_\mathrm{baseline})C \min_{i}\{R_i\}}
\\&=& \underbrace{\frac{C}{\mathbb{E}[b(\mathbf{s}_p)]}\frac{(1-\alpha_\mathrm{baseline})}{(1-\alpha_\mathrm{baseline}^{L+1}) } S \min_{i}\{R_i\}}_{\mathrm{Throughput~component}}  \underbrace{(1-\alpha_\mathrm{baseline}^{L+1}) \mathbb{E}[v(\mathbf{s}_p)]}_{\mathrm{Quality~component}}
\\&=& \frac{\mathbb{E}[v(\mathbf{s}_p)] C S(1-\alpha_\mathrm{baseline}) \min_{i}\{R_i\}}{\mathbb{E}[b(\mathbf{s}_p)]}.  \label{eq:PT-baseline2}
\end{eqnarray}

\noindent We make the following two key observations regarding the result in \eqref{eq:PT-baseline2}:

\begin{enumerate}
\item In the absence of packet prioritization, unequal modulation is not beneficial. This is because the throughput is limited by the worst spatial stream as evident by the term $\min_{i}\{R_i\}$.
\item In the absence of packet prioritization, the objective does not depend on the retransmission limit $r$. This is due to the fact that the loss in throughput due to retransmission is compensated by a gain in video quality and vice versa.
\end{enumerate}

Therefore, for the baseline case, we consider the same modulation order $M$ for all streams. Further, we select the modulation order $M$ and coding rate $C$ to maximize the post retransmission throughput, that is,

\begin{eqnarray}
\{M^*,C^*\} &=& \mathrm{argmax}_{M\in\mathcal{M},C\in\mathcal{C}}\left\{C R \frac{1-\alpha_\mathrm{baseline}(\boldsymbol{\gamma},M,C)}{1-\alpha_\mathrm{baseline}(\boldsymbol{\gamma},M,C)^{L+1}}\right\} .
\end{eqnarray}

\noindent where $R = B \log_2(M)$. Now, we write the gain $G = \mathbb{E}_\mathbf{H}\left[{WT}_{\textrm{prioritized}}\right]/\mathbb{E}_\mathbf{H}\left[{WT}_{\textrm{baseline}}\right]$ as follows

\begin{eqnarray}
G &=& \underbrace{\frac{\mathbb{E}_\mathbf{H}[\sum_{i=1}^{S}{(1-\alpha_i^{L + 1}) \int_{\hat{v}_{i}}^{\hat{v}_{i+1}}{v f_v(v) \mathrm{d}v}}]}{\mathbb{E}_\mathbf{H}[(1-\alpha_\mathrm{baseline}^{L+1}) \int_{0}^{1}{v f_v(v) \mathrm{d}v}]}}_{\textrm{Packet Prioritization Gain $G_\mathrm{PP}$}}\nonumber\times\underbrace{\frac{\mathbb{E}_\mathbf{H}[\max_{C}{\left\{C\sum_i{\max_{M_i}\{R_i/r_i\}}\right\}}]}{\mathbb{E}_\mathbf{H}[ S\max_{M,C}\{C R /r_\mathrm{baseline}\}]}}_{\textrm{Unequal Modulation Gain $G_\mathrm{UM}$}}\label{eq:G_baseline1}\\
&=& G_\mathrm{PP}\times G_\mathrm{UM}.
\end{eqnarray}

\noindent where $r_i = (1-\alpha_i^{L + 1})/(1-\alpha_i)$ and $r_\mathrm{baseline} = (1-\alpha_\mathrm{baseline}^{L + 1})/(1-\alpha_\mathrm{baseline})$ are the average number of retransmissions for the proposed and baseline scenarios respectively.



\subsection{Packet Prioritization Gain}

The first component of \eqref{eq:G_baseline1} is referred to as \emph{packet prioritization gain} and is expressed as follows

\begin{eqnarray}
G_\mathrm{PP}&=& \frac{\mathbb{E}_\mathbf{H}\left[\sum_{i=1}^{S}{(1-\alpha_i^{L + 1}) \mathbb{E}\left[v(\mathbf{s}_p)|\hat{v}_{i} \le v(\mathbf{s}_p) \le \hat{v}_{i+1} \right]}\right]}{\left(1-(1-\prod_{i=1}^{S}{\left(1-\mathbb{E}_\mathbf{H}\left[\alpha_i\right]\right)}^{1/S})^{L+1}\right) \mathbb{E}[v(\mathbf{s}_p)]}\label{eq:G_PP}.
\end{eqnarray}

\noindent It results from the fact that the more relevant packets are transmitted through the more reliable streams. Because streams are ordered by the post-retransmission success probability $1-\alpha_i^{L+ 1}$, the packet prioritization gain is always greater than 1. We note that this gain is highest when both the packet loss visibility and the per-stream SNRs exhibit high variability. Furthermore, if infinite retransmissions are allowed, this gain converges to one since all packets are eventually received successfully. The dependence on $\mathbf{H}$ in \eqref{eq:G_PP} is through both the loss visibility thresholds $\{\hat{v}_{i}\}_{i=1}^{S}$ and the PERs $\{\alpha_{i}\}_{i=1}^{S}$. \emph{The packet prioritization gain represents a reduction in loss visibility, i.e., a video quality gain}.

\subsection{Unequal Modulation Gain}

The second component of \eqref{eq:G_baseline1} is referred to as the \emph{unequal modulation gain} and is expressed as follows

\begin{eqnarray}
G_\mathrm{UM} &=& \frac{\mathbb{E}_\mathbf{H}[\max_{C}{\left\{C\sum_i{\max_{M_i}\{R_i/r_i\}}\right\}}]}{\mathbb{E}_\mathbf{H}[ S\max_{M,C}\{C R /r_\mathrm{baseline}\}]}\label{eq:G_UM}.
\end{eqnarray}

\noindent It corresponds to the throughput averaged over spatial streams divided by the throughput on the worst spatial stream. It results from the fact that the optimal transmission policy can opportunistically increase the rate on the stronger streams to enable low latency delivery of high priority packets. Conversely, in conventional MIMO transmission with a fixed modulation order, the performance achieved is limited by the performance on the worst stream. This justifies why the unequal modulation gain is the achievable throughput averaged over all streams divided by the achievable throughput on the worst stream. The dependence on $\mathbf{H}$ in \eqref{eq:G_UM} is through the PERs $\{\alpha_{i}\}_{i=1}^{S}$ which impact the per-stream throughputs $\{R_i\}_{i=1}^{S}$ and the retransmission overhead $\{r_{i}\}_{i=1}^{S}$. \emph{The unequal modulation gain results in an increase in throughput.}

\subsection{Impact of Limited Feedback}

The expressions for ${WT}_\mathrm{prioritized}$ and ${WT}_\mathrm{baseline}$ are in terms of the error probability $p^\mathrm{success}_i$, which in turn depends on the post processing SNR vector $\boldsymbol{\gamma} = \{\gamma_i (\mathbf{H})\}_{i=1}^{S}$. Thus, they apply equivalently under limited feedback given that $\gamma_i (\mathbf{H})$ is computed using \eqref{eq:SNR} according to the selected precoder. We compute the gains under limited feedback by taking the expectation of the individual gains for each channel state given its mapping to the corresponding codeword. This corresponds to $G = \mathbb{E}_{\mathcal{F}}\left[{WT}_{\textrm{prioritized}}\right]/\mathbb{E}_{\mathcal{F}}\left[{WT}_{\textrm{baseline}}\right]$, i.e.,

\begin{eqnarray}
G &=& \underbrace{\frac{\mathbb{E}_\mathbf{\mathcal{F}}[\sum_{i=1}^{S}{(1-\alpha_i^{L + 1}) \int_{\hat{v}_{i}}^{\hat{v}_{i+1}}{v f_v(v) \mathrm{d}v}}]}{\mathbb{E}_\mathbf{\mathcal{F}}[(1-\alpha_\mathrm{baseline}^{L+1}) \int_{0}^{1}{v f_v(v) \mathrm{d}v}]}}_{\textrm{Packet Prioritization Gain $G_\mathrm{PP}$}}\nonumber\times\underbrace{\frac{\mathbb{E}_\mathbf{\mathcal{F}}[\max_{C}{\left\{C\sum_i{\max_{M_i}\{R_i/r_i\}}\right\}}]}{\mathbb{E}_\mathbf{\mathcal{F}}[ S\max_{M,C}\{C R /r_\mathrm{baseline}\}]}}_{\textrm{Unequal Modulation Gain $G_\mathrm{UM}$}}\label{eq:G_LF}\\
&=& G_\mathrm{PP}\times G_\mathrm{UM}.
\end{eqnarray}

\section{Results and Analysis}\label{results}

In this section, we first evaluate the proposed loss visibility based MIMO transmission policies using H.264 encoded bit streams under different antenna configurations. Next, we present numerical results to quantify the packet prioritization and unequal modulation gains. Each entry of the channel matrix corresponds to a flat Rayleigh fading channel. The system bandwidth is 1 MHz. The set of possible M-QAM constellations is $\mathcal{M} = \{2,4,16,64\}$ corresponding to BPSK, 4-QAM, 16-QAM, and 64-QAM. The set of possible coding rates is $\mathcal{C} = \{1/2,2/3,3/4,5/6\}$. 

\subsection{Video Quality Gains on H.264 Sequences}

\begin{figure}[t!]
  \centering
  \includegraphics[width=0.92\linewidth]{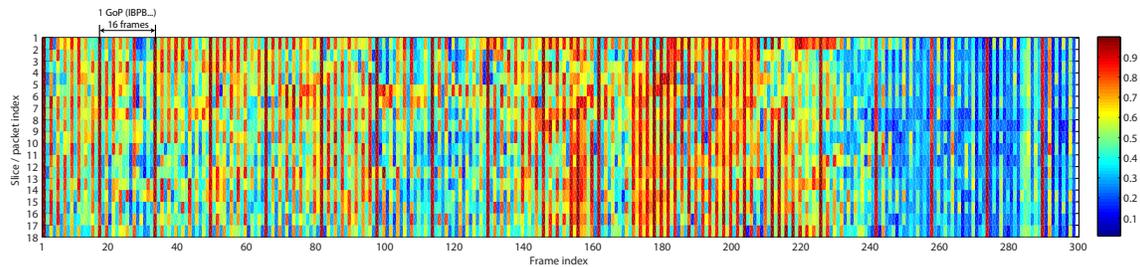}
\caption{Loss visibility map of the Foreman video sequence encoded with H.264/AVC using a IBPBP GoP structure with 18 horizontal slices per frame and a GoP duration of 16.\label{fig:LV_map}}
\end{figure}

To evaluate the video quality gain from the loss visibility based prioritization policy, we test the proposed algorithm on the Foreman video sequence \cite{videoSequences} encoded with H.264/AVC. The GoP structure is $IBPBP\cdots$ and the GoP duration is 16 frames. The MB size is $16\times16$ and we use the  CIF resolution of $352 \times 288$. The video frame is divided into horizontal slices where each slice is 22 MBs wide and 1 MB high. Thus, each frame corresponds to 18 slices and each slice is transmitted as one packet. The decoder uses motion copy error concealment. Loss visibility estimation is applied based on \cite{LinPrioritization} as described in \S\ref{LV-estimation}. Figure \ref{fig:LV_map} shows the resulting loss visibility scores for each frame/slice for the Foreman video sequence. Several observations are in order.

\begin{enumerate}
\item \emph{Frame type:} The variability of the visibility across frames is clear. For instance, the $I$ frames can be noticed as dark red every GoP interval. Furthermore, the odd-numbered frames corresponding to $P$ have higher loss visibility than the even-numbered $B$ frames.
\item \emph{Subject/background motion:} Face motion between Frame 1 and Frame 170 cause high loss visibility for some slices depending on the spatial location of motion. Background motion between Frame 170 and Frame 220 contributes an overall increase in loss visibility. Beyond that, the lack of object and background motion causes an overall drop in loss visibility.
\item \emph{Error propagation:} For odd-numbered $P$ frames, it can be noticed that the packet loss visibility captures the severity of potential error propagation by decaying for $P$ frames towards the end of the GoP, i.e., close to the next reference frame.
\end{enumerate}

\begin{figure}[h!]
  \centering
\subfigure[Packet-Stream Mapping]{\label{fig:case-study1}\includegraphics[width=\linewidth]{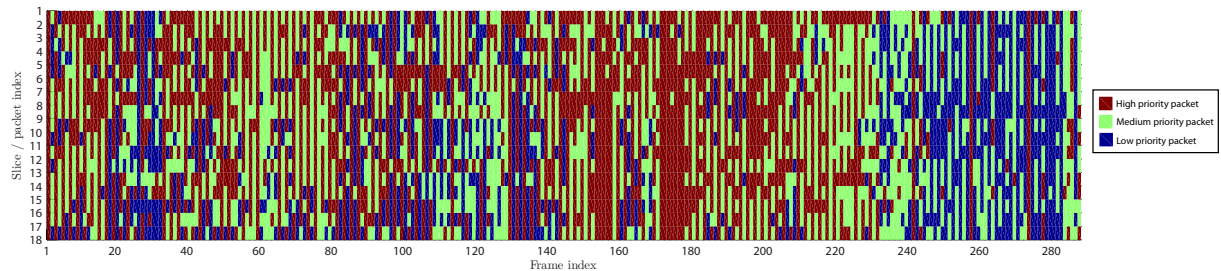}}\\
\subfigure[Received frame 223 with and without prioritization ]{\label{fig:case-study2} \includegraphics[height=0.19\linewidth]{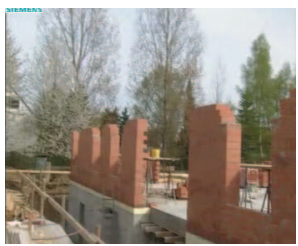}
\hspace{0.2cm}  \includegraphics[height=0.19\linewidth]{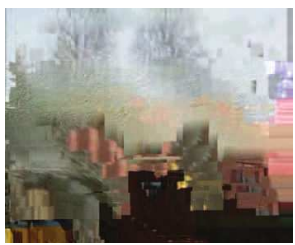}
}
\subfigure[Comparison of video quality of the received videos]{\label{fig:case-study3} \includegraphics[height=0.19\linewidth]{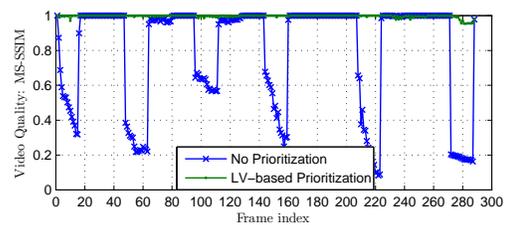}}
  \caption{Case study of the loss visibility-based prioritization policy for the Foreman video sequence with $4\times4$ MIMO system, $S = 3$ streams, and $E_s/N_0 = 5$ dB. The retransmission limit is $r = 4$.\label{fig:case-study}}
\end{figure}

Figure \ref{fig:case-study} applies the loss visibility based prioritization policy to the Foreman video sequence \cite{videoSequences} for a $4\times4$ MIMO system, $S = 3$ streams/classes, and $E_s/N_0 = 5$ dB. The retransmission limit is $r = 4$ and the channel coherence time is equal to 1 GoP corresponding to a low mobility environment. Figure \ref{fig:case-study1} shows the mapping of each video packet to the corresponding spatial stream. Packets mapped to the best spatial stream are referred to as high priority packets and vice versa. The corresponding video quality is shown in Figure \ref{fig:case-study3} in comparison with the baseline, whereby the symbols corresponding to each packet are mapped to all spatial streams, for the same channel realization. \emph{Despite having 460 packet losses post-retransmission, the mean video quality with prioritization is 0.997 on the MS-SSIM scale whereas the mean video quality without prioritization is 0.802.} With packet prioritization, losses affect only packets where error concealment can conceal the loss from being visible to the average viewer. In contrast, the error propagation effect is very severe in the case of no prioritization. The received and concealed frames with index 223 of the Foreman sequence are shown in Figure \ref{fig:case-study2} to further demonstrate the difference in video quality.

\begin{figure}[t!]
  \centering
  \includegraphics[height=0.35\linewidth]{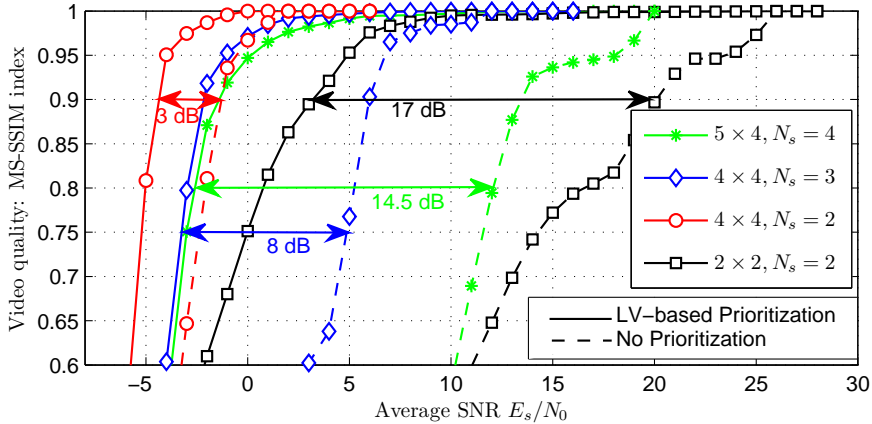}
  \caption{Comparison of the loss visibility-based packet prioritization vs. non-prioritized MIMO precoding for H.264-encoded Foreman sequence for different antenna configurations over a range of SNRs. The retransmission limit is $r = 4$ and the channel coherence time is 1 GoP.\label{fig:Quality}}
\end{figure}

Figure \ref{fig:Quality} demonstrates the video quality gains for a range of antenna configurations for the Foreman video sequence encoded with the same properties as previously described. The video quality at each data point is the frame-averaged quality further averaged over 10 different channel realizations. The same channel realizations are used for the two cases. The first observed trend is that for a fixed antenna configuration, the gains are maximized when $S = \mathrm{min}(N_t,N_r)$. This is because the large variability in the post-processing SNRs across streams enables more effective packet prioritization. Furthermore, increasing the number of antennas for a fixed number of streams improves video quality but reduces the video quality gain. The maximum gain is reported for a $2\times2$ setting where a video quality of 0.9 requires $E_s/N_0 = 3$ dB with prioritization versus $E_s/N_0 = 20$ dB without prioritization. Furthermore, gains in the excess of 10 dB are achieved over a range of antenna configurations.

\subsection{Throughput Gains}

\begin{figure*}[t!]
  \centering
  \subfigure[Different antenna configurations with $S=2$ streams]{\label{fig:Gain_LB_retx} \includegraphics[height=0.36\linewidth]{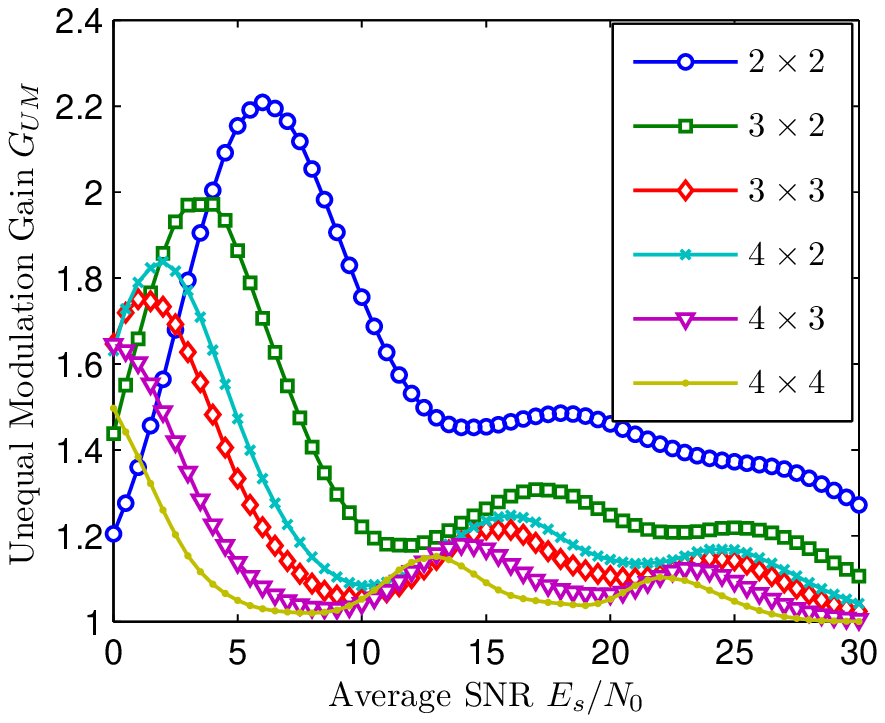}}
  \subfigure[Different number of streams for a $4\times4$ system]{\label{fig:Gain_LB_4x4} \includegraphics[height=0.36\linewidth]{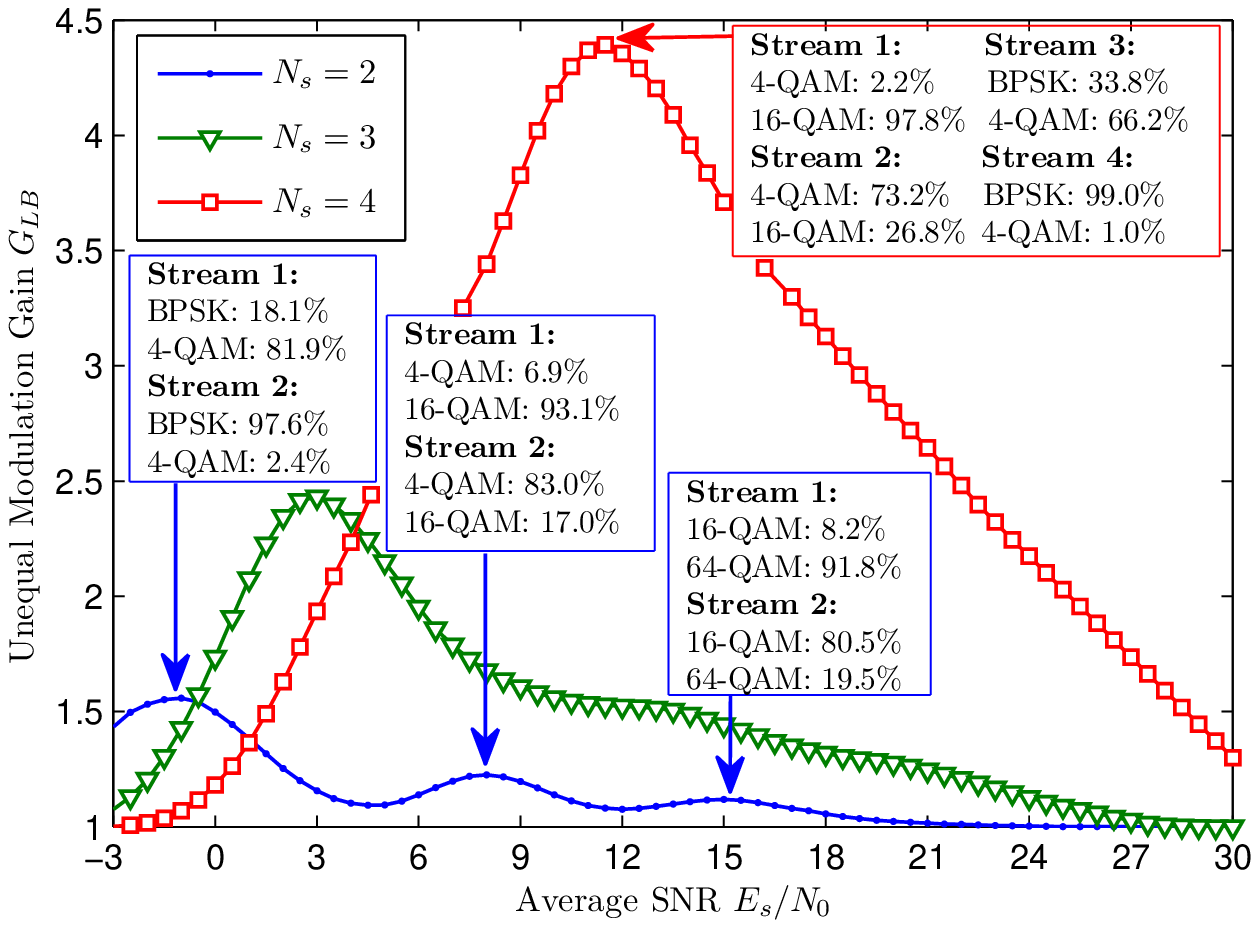}}
  \caption{Analysis of the unequal modulation gain $G_{UM}$. It corresponds to the throughput averaged over spatial streams divided by the throughput on the worst spatial stream. The peaks correspond to operating points where the modulation orders across streams are very likely to be ``unequal''. \label{fig:Gain_LB}}
\end{figure*}

Having shown that significant video quality gains are achieved by the loss visibility-based video transmission policies, we then examine the throughput gains by plotting the closed-form unequal modulation gain expression derived in \S\ref{gains}. Recall that the throughput gain is achieved due to the ability to leverage unequal modulation on the stronger spatial streams.

In Figure \ref{fig:Gain_LB}, we examine the unequal modulation gain $G_\mathrm{UM}$, defined in \eqref{eq:G_baseline1}. Figure \ref{fig:Gain_LB_retx} shows the gain for $S = 2$ spatial streams with different antenna configurations. Recall from the unequal modulation gain expression that the gain is maximized when the per-stream throughputs exhibit the highest variability among streams. In a two stream setup, this corresponds to the case where the difference between the throughput on the two steams is maximal. Thus, for $S = 2$, a $2 \times 2$ system gains more than a $4 \times 4$ system. In a $4 \times 4$ system with $S = 2$, the diversity and channel hardening reduce the gains from the proposed prioritization policy because the supported modulation orders per stream are equivalent for most channel realizations and the achievable throughput on the two streams is comparable. In Figure \ref{fig:Gain_LB_4x4}, we plot the unequal modulation gain for a $4 \times 4$ system for different numbers of spatial streams $S$. In the medium to high SNR regime, for the same $N_\mathrm{t} \times N_\mathrm{r}$ configuration, more streams provide higher gains versus non-video aware approaches since the condition number of the effective channel $\mathbf{H}\mathbf{F}_S$ is likely to be higher making it possible for video-aware techniques to make use of  the diverse channel statistics among streams. For $S = 2$ and $S = 4$, we show the fractional use of each modulation scheme at the peak operating points. For $S = 2$ at $E_\mathrm{s}/N_0 = -1$ dB, the best stream can support 4-QAM for most realizations while the worst stream can only support BPSK. A similar observation follows at 8 dB and 15 dB for 16-QAM and 64-QAM. Conversely, at 4 dB (resp. 12 dB), both streams support 4-QAM (resp. 16-QAM) for most channel realizations. Thus, the gain $G_\mathrm{UM}$ is close to 1.

\begin{figure*}[t!]
  \centering
  \subfigure[Loss visibility-based packet prioritization]{\label{fig:PT_avg1} \includegraphics[width=0.48\linewidth]{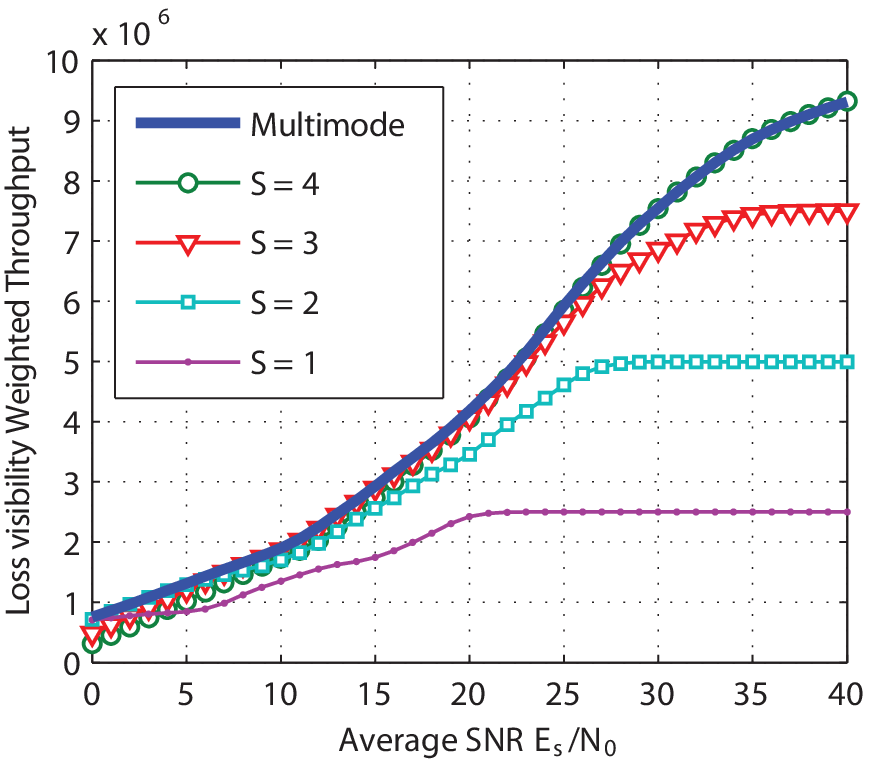}}
  \subfigure[Non-prioritized MIMO precoding]{\label{fig:PT_avg2} \includegraphics[width=0.48\linewidth]{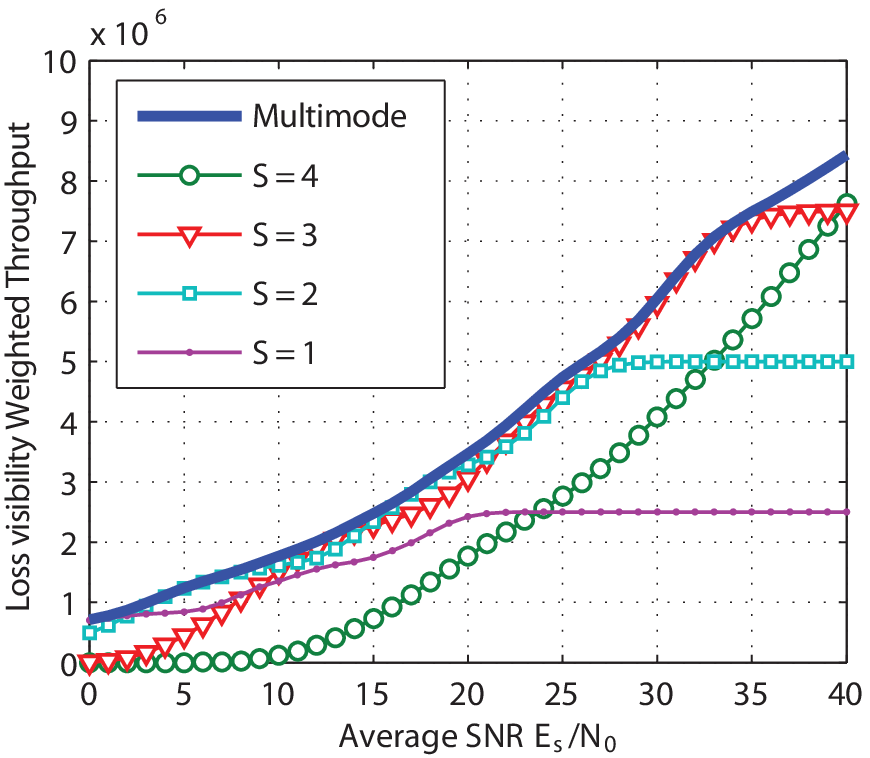}}
  \caption{Comparison of the loss visibility weighted throughput objective achieved by loss visibility-based packet prioritization and non-prioritized MIMO transmission for a $4\times4$ MIMO system for different number of streams.\label{fig:PT_avg}}
\end{figure*}

In Figure \ref{fig:PT_avg}, we plot the weighted throughput objective achieved by loss visibility-based packet prioritization vs. non-prioritized MIMO precoding for a $4\times4$ MIMO system under different numbers of spatial streams. For beamforming ($S=1$), the performance is equivalent since there is only a single packet class. Comparing Figures \ref{fig:PT_avg1} and \ref{fig:PT_avg2} for $S>1$, we clearly observe that for the same SNR, the objective achieved with packet prioritization is higher. Even comparing multimode prioritized transmission with multimode non-prioritized transmission where gains are expected to drop, we notice a 3 dB gain in the low SNR regime and a 6 dB gain in the high SNR regime.



\subsection{Prioritized Transmission with Limited Feedback}

\begin{figure}[t!]
  \centering
   \includegraphics[width=0.55\linewidth]{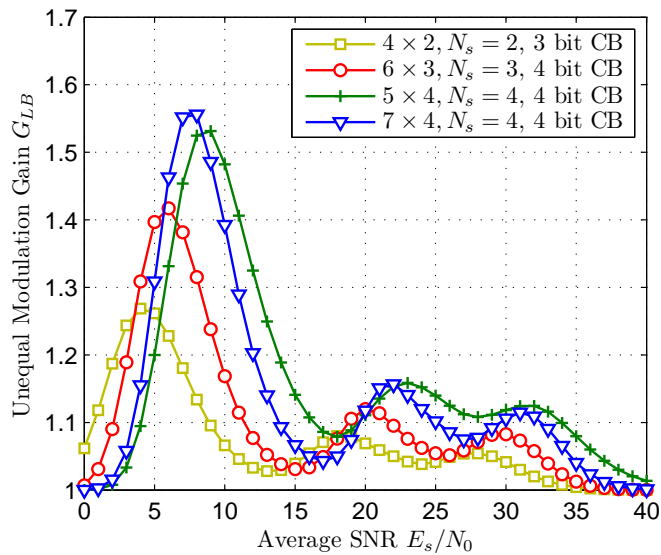}
  \caption{Unequal modulation gain achieved with limited feedback for different codebook sizes and antenna configurations.\label{fig:Gain_LF}}
\end{figure}

Figure \ref{fig:Gain_LF} shows the unequal modulation gain with limited feedback for different codebook sizes and antenna configurations. The codebooks are obtained using Grassmannian subspace packing with the chordal subspace distance measure \cite{Grass}. As expected, the gains increase as the codebook size increases as well as for larger number of spatial streams. With only 2 spatial streams in a $4\times 2$ antenna configuration and a 3 bit codebook, $27\%$ throughput increase is achieved. With 4 spatial streams in a $7\times 4$ antenna configuration and a 4 bit codebook, $56\%$ throughput increase is achieved. The trends of the gains closely follows those in Figure \ref{fig:Gain_LB} corresponding to perfect CSI feedback. In terms of the nominal gain values, we observe that with codebook-based limited feedback, the gain drops because the unequal stream quality cannot be fully utilized due to channel quantization errors. Such errors cause the gap between the post processing SNRs on the best and worst stream to tighten, thus reducing the achievable gain.

\begin{figure}[t!]
  \centering
  \includegraphics[height=0.55\linewidth]{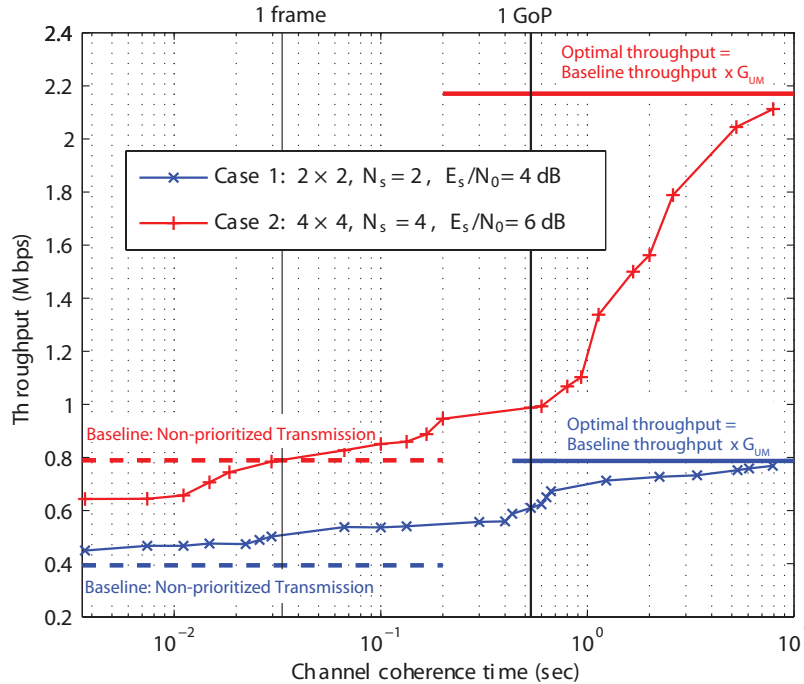}
  \caption{Analysis of the effect of channel coherence time on the achievability of the throughput gain for the Foreman video sequence with 2 different antenna configurations. \label{fig:Channel_coh}}
\end{figure}

\subsection{Impact of Mobility}

Although the analysis applies to any channel coherence time larger than one packet, the underlying assumption in the proof of Theorem 1 is that the packets observed within a channel coherence time are representative of the loss visibility distribution. Otherwise, the observed short-term loss visibility distribution will be different from the distribution estimated using kernel density estimation causing a \emph{loss visibility distribution mismatch}. It follows that the gains in Figure \ref{fig:Gain_LB} are an upper bound that apply with a fairly large channel coherence time. For a more realistic analysis of the throughput gain, we simulate the proposed algorithm in Figure \ref{fig:Channel_coh} with a variable channel coherence time ranging from $S$ packets to several GoPs under 2 antenna configurations. For a $2\times2$ system, the throughput always exceeds that of the baseline but the theoretical 2x load balancing gain reported in Figure \ref{fig:Gain_LB} is only achieved if the channel is fairly static for few seconds. For a practical low mobility setup where the channel coherence is equal to one GoP, 1.5x out of the theoretical 2x gain is achieved. For a $4\times4$ MIMO system, the throughput exceeds that of the baseline when the channel is at least $35$ ms equivalent to one video frame. Beyond that, for a channel coherence of one GoP, 1.25x throughput gain is achieved.

\section{Conclusion}\label{conclusion}

We proposed a cross-layer architecture for prioritized packet delivery over a MIMO PHY layer based on loss visibility taking advantage of the large variability in loss visibility due to the video source and encoder features. We presented a loss visibility-based thresholding policy that maps different packets to different spatial streams and derived the optimal thresholding policy for any loss visibility distribution. The proposed architecture requires minimal additional cross-layer overhead while achieving quality and capacity gains. We demonstrated gains in the excess of 10 dB with different antenna configurations on H.264 encoded video sequences.

\newpage

\appendices

\section{Proof of Lemma 1}

\noindent \textbf{Lemma 1. }The optimal packet-stream mapping is such that $\mathcal{V}_{i}$ has the form $\mathcal{V}_{i}=[\hat{v}_i,\hat{v}_{i+1}]$ where $\cup_{i=1}^{S} \mathcal{V}_{i} = [0,1]$. Furthermore, for any two packets $\mathbf{s}_1$ and $\mathbf{s}_2$ s.t. $v(\mathbf{s}_1)<v(\mathbf{s}_2)$, $\mathbf{s}_1\in \mathcal{V}_i$ and $\mathbf{s}_2\in \mathcal{V}_{k}$ where $p^\mathrm{success}_i\le p^\mathrm{success}_{k}$. It follows that the streams should be ordered by the probability of success $p^\mathrm{success}_i \le p^\mathrm{success}_{i+1}$.

\textbf{Proof:} Consider two video packets $\mathbf{s}_1$ and $\mathbf{s}_2$ such that $v(\mathbf{s}_1)<v(\mathbf{s}_2)$. Assume the packet-stream mapping is such that $\mathbf{s}_2\in \mathcal{V}_i'$ and $\mathbf{s}_1\in \mathcal{V}_k'$ where $p^\mathrm{success}_i = 1-\alpha_i^{L + 1} \le 1-\alpha_{k}^{L + 1} = p^\mathrm{success}_{k}$. We switch the mapping of packets $\mathbf{s}_1$ and $\mathbf{s}_2$, that is, $\mathcal{V}_i = \mathcal{V}_i' + \{\mathbf{s}_2\} - \{\mathbf{s}_1\}$ and $\mathcal{V}_k = \mathcal{V}_k' + \{\mathbf{s}_1\} - \{\mathbf{s}_2\}$. We show that the corresponding objective function $WT' \le WT$.

\begin{eqnarray}
{WT}' &=& \frac{(\sum_{l\notin \{i,k\}}{p^\mathrm{success}_l~ \sum_{\mathbf{s}\in\mathcal{V}_l}{ v(\mathbf{s})  }}) + p^\mathrm{success}_i~ \sum_{\mathbf{s}\in\mathcal{V}_i'}{ v(\mathbf{s})  } + p^\mathrm{success}_k~ \sum_{\mathbf{s}\in\mathcal{V}_k'}{ v(\mathbf{s})  }  }{\max\{\max_{l\notin \{i,k\}} {t_l(\gamma_l,M_l,C,\mathcal{V}_l)}, t_i(\gamma_i,M_i,C,\mathcal{V}_i'),t_k(\gamma_k,M_k,C,\mathcal{V}_k')\}}\\
&=& \frac{(\sum_{l\notin \{i,k\}}{p^\mathrm{success}_l~ \sum_{\mathbf{s}\in\mathcal{V}_l}{ v(\mathbf{s})  }}) + p^\mathrm{success}_i~ \sum_{\mathbf{s}\in\mathcal{V}_i}{ v(\mathbf{s})  } + p^\mathrm{success}_k~ \sum_{\mathbf{s}\in\mathcal{V}_k}{ v(\mathbf{s})  } }{\max\{\max_{l\notin \{i,k\}} {t_l(\gamma_l,M_l,C,\mathcal{V}_l)}, t_i(\gamma_i,M_i,C,\mathcal{V}_i'),t_k(\gamma_k,M_k,C,\mathcal{V}_k')\}}\nonumber\\
&& +\frac{(p^\mathrm{success}_i - p^\mathrm{success}_k) (v(\mathbf{s}_2) - v(\mathbf{s}_1))}{\max\{\max_{l\notin \{i,k\}} {t_l(\gamma_l,M_l,C,\mathcal{V}_l)}, t_i(\gamma_i,M_i,C,\mathcal{V}_i'),t_k(\gamma_k,M_k,C,\mathcal{V}_k')\}}\\
&<& \frac{\sum_{l}{p^\mathrm{success}_l~ \sum_{\mathbf{s}\in\mathcal{V}_l}{ v(\mathbf{s})  }}   }{\max\{\max_{l\notin \{i,k\}} {t_l(\gamma_l,M_l,C,\mathcal{V}_l)}, t_i(\gamma_i,M_i,C,\mathcal{V}_i'),t_k(\gamma_k,M_k,C,\mathcal{V}_k')\}}
\label{eq:step1} \\
&\le& \frac{\sum_{l}{p^\mathrm{success}_l~ \sum_{\mathbf{s}\in\mathcal{V}_l}{ v(\mathbf{s})  }}  }{\max\{\max_{l\notin \{i,k\}} {t_l(\gamma_l,M_l,C,\mathcal{V}_l)}, t_i(\gamma_i,M_i,C,\mathcal{V}_i),t_k(\gamma_k,M_k,C,\mathcal{V}_k)\}}
\label{eq:step2} \\
&=& \frac{\sum_{l}{p^\mathrm{success}_l~ \sum_{\mathbf{s}\in\mathcal{V}_l}{ v(\mathbf{s})  }}}{\max_l t_l} = {WT}
\end{eqnarray}

\noindent where \eqref{eq:step1} follows because $v(\mathbf{s}_1)<v(\mathbf{s}_2)$ and $p^\mathrm{success}_i \le p^\mathrm{success}_{k}$ by definition and $p^\mathrm{success}_{i} \le p^\mathrm{success}_{k}$ by the proposed ordering. Next, we show the transition to \eqref{eq:step2} by showing it separately in the following four possible cases. For brevity, we denote by $\mathrm{Thr}(i)$ the throughput on the $i^{\mathrm{th}}$ stream in the derivation below.

\begin{enumerate}
      \item $\{\mathrm{argmax}(t_l,t_i(\mathcal{V}_i'),t_k(\mathcal{V}_k'))) = i$, $\mathrm{argmax}(t_l,t_i(\mathcal{V}_i),t_k(\mathcal{V}_k)) = i\}$: In this case, switching the ordering improves the objective since $v(\mathbf{s}_1)<v(\mathbf{s}_2)$ and $b(\mathbf{s}_1)<b(\mathbf{s}_2)$, thus $t_i(\mathcal{V}_i) < t_i(\mathcal{V}_i')$.
\item $\{\mathrm{argmax}(t_l,t_i(\mathcal{V}_i'),t_k(\mathcal{V}_k'))) = k$, $\mathrm{argmax}(t_l,t_i(\mathcal{V}_i),t_k(\mathcal{V}_k)) = k\}$: While this reduces the objective since $t_k(\mathcal{V}_k)>t_k(\mathcal{V}_k)$, we show by contradiction that it never occurs. We have $\mathrm{argmax}(t_l,t_i(\mathcal{V}_i'),t_k(\mathcal{V}_k'))) = k \Longrightarrow b(\mathbf{s}_2)/\mathrm{Thr}(i)<b(\mathbf{s}_1)/\mathrm{Thr}(k)$ and $\mathrm{argmax}(t_l,t_i(\mathcal{V}_i),$ $t_k(\mathcal{V}_k))) = k \Longrightarrow b(\mathbf{s}_2)/\mathrm{Thr}(k)>b(\mathbf{s}_1)/\mathrm{Thr}(i)$. Thus, $\mathrm{Thr}(i)/\mathrm{Thr}(k) < b(\mathbf{s}_2)/b(\mathbf{s}_1) < \mathrm{Thr}(k)/\mathrm{Thr}(i)$. Since $b(\mathbf{s}_2)/b(\mathbf{s}_1)>1$, we have $1 < \mathrm{Thr}(k)/\mathrm{Thr}(i) \Longrightarrow \mathrm{Thr}(i)<\mathrm{Thr}(k) \Longrightarrow b(\mathbf{s}_1)/\mathrm{Thr}(i)>b(\mathbf{s}_1)/\mathrm{Thr}(k)$. Combining with $b(\mathbf{s}_2)/\mathrm{Thr}(i)<b(\mathbf{s}_1)/\mathrm{Thr}(k)$, we obtain $b(\mathbf{s}_2)/\mathrm{Thr}(i)< b(\mathbf{s}_1)/\mathrm{Thr}(i) \Longrightarrow b(\mathbf{s}_2)<b(\mathbf{s}_1)$. Thus, we have a contradiction.
\item $\{\mathrm{argmax}(t_l,t_i(\mathcal{V}_i'),t_k(\mathcal{V}_k'))) = i$, $\mathrm{argmax}(t_l,t_i(\mathcal{V}_i),t_k(\mathcal{V}_k)) = k\}$: We show by contradiction that this case never occurs. $\mathrm{argmax}(t_l,t_i(\mathcal{V}_i'),t_k(\mathcal{V}_k'))) = i \Longrightarrow b(\mathbf{s}_2)/\mathrm{Thr}(i)>b(\mathbf{s}_1)/\mathrm{Thr}(k)$ and $\mathrm{argmax}(t_l,t_i(\mathcal{V}_i),t_k(\mathcal{V}_k)) = k \Longrightarrow b(\mathbf{s}_2)/\mathrm{Thr}(k)>b(\mathbf{s}_1)/\mathrm{Thr}(i)$. Thus, $b(\mathbf{s}_2)/b(\mathbf{s}_1) > \mathrm{Thr}(k)/\mathrm{Thr}(i)$ and $b(\mathbf{s}_2)/b(\mathbf{s}_1) > \mathrm{Thr}(i)/\mathrm{Thr}(k)$ which is a contradiction.
\item $\{\mathrm{argmax}(t_l,t_i(\mathcal{V}_i'),t_k(\mathcal{V}_k'))) = k$, $\mathrm{argmax}(t_l,t_i(\mathcal{V}_i),t_k(\mathcal{V}_k)) = i\}$: We show by contradiction that this case never occurs. $\mathrm{argmax}(t_l,t_i(\mathcal{V}_i'),t_k(\mathcal{V}_k'))) = k \Longrightarrow b(\mathbf{s}_2)/\mathrm{Thr}(i)<b(\mathbf{s}_1)/\mathrm{Thr}(k)$ and $\mathrm{argmax}(t_l,t_i(\mathcal{V}_i),t_k(\mathcal{V}_k)) = i \Longrightarrow b(\mathbf{s}_2)/\mathrm{Thr}(k)<b(\mathbf{s}_1)/\mathrm{Thr}(i)$. Thus, $b(\mathbf{s}_2)/b(\mathbf{s}_1) < \mathrm{Thr}(k)/\mathrm{Thr}(i)$ and $b(\mathbf{s}_2)/b(\mathbf{s}_1) < \mathrm{Thr}(i)/\mathrm{Thr}(k)$ which is a contradiction.
    \end{enumerate}

Thus, the proposed ordering maximizes the objective function and $\hat{v}_i$ is the threshold between ordered spatial stream $i-1$ and $i$.

\section{Proof of Lemma 2}

\noindent \textbf{Lemma 2. }\emph{If the streams are ordered by the post-retransmission success probability, i.e., $p^\mathrm{success}_i \le p^\mathrm{success}_{i+1}~\forall i = 1,\cdots, N_s - 1$, then the gradient $\partial {WT}/\partial \hat{v}_i$ satisfies the following properties:}

\begin{enumerate}
\item $\partial {WT}/\partial \hat{v}_{\tilde{i}}\ge0$ where $\tilde{i} = \mathrm{argmax}~t_i$
\item $\partial {WT}/\partial \hat{v}_i\le0~\forall i\ne \tilde{i}$
\end{enumerate}

\textbf{Proof:} First, the gradient of ${WT}(\hat{\mathbf{v}},\mathbf{M},C,S)$ with respect to $\hat{v}_i$ is $\partial {WT}/\partial \hat{v}_i = (h\partial g/\partial \hat{v}_i - g\partial h/\partial \hat{v}_i)/h^2$ where $g = \left[\sum_{i=1}^{S}{(1-\alpha_i^{L + 1}) \int_{\hat{v}_{i}}^{\hat{v}_{i+1}}{v f_v(v) \mathrm{d}v}}\right]$ and $h = \mathbb{E}[b(\mathbf{s}_p)](F_v(\hat{v}_{\tilde{i}+1})-F_v(\hat{v}_{\tilde{i}}))(1-\alpha_{\tilde{i}}^{L + 1})/C R_{\tilde{i}}(1-\alpha_{\tilde{i}})$ are the numerator and denominator of \eqref{eq:T_R_S1}. The components of the gradient are

\vspace{-0.5cm}

\begin{equation}
\frac{\partial g}{\partial \hat{v}_i} = ( \alpha_i^{L+1}- \alpha_{i-1}^{L+1}) \hat{v}_{i} f_v(\hat{v}_{i})
\end{equation}

\noindent where we used the fact that $\partial (\int_{0}^{\hat{v}_{1}}{v f_v(v) \mathrm{d}v}) /\partial \hat{v}_1 = \lim_{\epsilon\rightarrow 0} (\int_{\hat{v}_{1}}^{\hat{v}_{1}+\epsilon}{v f_v(v) \mathrm{d}v}/\epsilon) = \hat{v}_{1} f_v(\hat{v}_{1})$. Furthermore, the gradient corresponding to $h$ is

\begin{equation}
\frac{\partial h}{\partial \hat{v}_i} = \left\{ \begin{array}{ll} \mathbb{E}[b(\mathbf{s}_p)] f_v(\hat{v}_{i})(1-\alpha_{i-1}^{L + 1})/(C R_{i-1} (1-\alpha_{i-1})) &\hspace{-0.2cm}\textrm{if $i=\tilde{i}+1$}\\
- \mathbb{E}[b(\mathbf{s}_p)] f_v(\hat{v}_{i})(1-\alpha_i^{L + 1})/(C R_{i} (1-\alpha_i)) & \hspace{-0.2cm}\textrm{if $i=\tilde{i}$}\\
0 & \hspace{-0.2cm}\mathrm{otherwise}.\end{array}\right.
\end{equation}

\noindent Next, we prove part 1 of the Lemma. From the expressions for $\partial g/\partial \hat{v}_i$ and $\partial h/\partial \hat{v}_i$, it follows that

\begin{eqnarray}
\hspace{-0.5cm} \frac{\partial {WT}}{\partial \hat{v}_{\tilde{i}}}h^2 \hspace{-0.3cm}&=&\hspace{-0.3cm} (\alpha_{\tilde{i}}^{L+1}- \alpha_{{\tilde{i}}-1}^{L+1}) \hat{v}_{i} f_v(\hat{v}_{i})\frac{\mathbb{E}[b(\mathbf{s}_p)](F_v(\hat{v}_{\tilde{i}+1})-F_v(\hat{v}_{\tilde{i}}))(1-\alpha_{\tilde{i}}^{L + 1})}{C R_{\tilde{i}}(1-\alpha_{\tilde{i}})} \nonumber\\&&+ \left(\sum_{i=1}^{S}{(1-\alpha_i^{L + 1}) \int_{\hat{v}_{i}}^{\hat{v}_{i+1}}{v f_v(v) \mathrm{d}v}}\right)\frac{\mathbb{E}[b(\mathbf{s}_p)] f_v(\hat{v}_{\tilde{i}})(1-\alpha_{\tilde{i}}^{L + 1})}{C R_{\tilde{i}} (1-\alpha_{\tilde{i}})}\label{eq:reduction1}
\\ &=&\hspace{-0.3cm} \frac{\mathbb{E}[b(\mathbf{s}_p)] f_v(\hat{v}_{\tilde{i}})(1-\alpha_{\tilde{i}}^{L + 1})}{C R_{\tilde{i}} (1-\alpha_{\tilde{i}})}\times\nonumber\\&&\hspace{-0.3cm} \left[\left(\sum_{i=1}^{S}{(1-\alpha_i^{L + 1}) \int_{\hat{v}_{i}}^{\hat{v}_{i+1}}{\hspace{-0.5cm}v f_v(v) \mathrm{d}v}}\right) - (\alpha_{{\tilde{i}}-1}^{L+1}-\alpha_{\tilde{i}}^{L+1})\hat{v}_{i}(F_v(\hat{v}_{\tilde{i}+1})-F_v(\hat{v}_{\tilde{i}}) )\right]\label{eq:reduction2}
\\ &\ge& \hspace{-0.3cm}\frac{\mathbb{E}[b(\mathbf{s}_p)] f_v(\hat{v}_{\tilde{i}})(1-\alpha_{\tilde{i}}^{L + 1})}{C R_{\tilde{i}} (1-\alpha_{\tilde{i}})}\times\nonumber\\&&\hspace{-0.3cm} \left[\left({(1-\alpha_{\tilde{i}}^{L + 1}) \int_{\hat{v}_{\tilde{i}}}^{\hat{v}_{{\tilde{i}}+1}}{\hspace{-0.5cm}v f_v(v) \mathrm{d}v}}\right) - ( \alpha_{{\tilde{i}}-1}^{L+1}-\alpha_{\tilde{i}}^{L+1})\hat{v}_{i}(F_v(\hat{v}_{\tilde{i}+1})-F_v(\hat{v}_{\tilde{i}}) )\right]\label{eq:reduction3}
\end{eqnarray}

\noindent where \eqref{eq:reduction3} follows because $\sum_{i=1}^{S}{ (1-\alpha_{{i}}^{L + 1})  V_i}\ge (1-\alpha_{\tilde{i}}^{L + 1})  V_{\tilde{i}}$. Next, using the fact that $\alpha_{{\tilde{i}}-1}^{L+1}\le 1$, we further reduce the expression to

\begin{eqnarray} \hspace{-0.7cm}\frac{\partial {WT}}{\partial \hat{v}_{\tilde{i}}}h^2 \hspace{-0.3cm}&\ge& \hspace{-0.3cm}\frac{\mathbb{E}[b(\mathbf{s}_p)] f_v(\hat{v}_{\tilde{i}})(1-\alpha_{\tilde{i}}^{L+ 1})}{C R_{\tilde{i}} (1-\alpha_{\tilde{i}})}\times\nonumber\\&&\hspace{-0.3cm} \left[\left({(1-\alpha_{\tilde{i}}^{L+ 1}) \int_{\hat{v}_{\tilde{i}}}^{\hat{v}_{{\tilde{i}}+1}}{\hspace{-0.5cm}v f_v(v) \mathrm{d}v}}\right) - ( 1-\alpha_{\tilde{i}}^{L+1})\hat{v}_{i}(F_v(\hat{v}_{\tilde{i}+1})-F_v(\hat{v}_{\tilde{i}}) )\right]\label{eq:reduction4}
\\&=& \hspace{-0.3cm}\frac{\mathbb{E}[b(\mathbf{s}_p)] f_v(\hat{v}_{\tilde{i}})(1-\alpha_{\tilde{i}}^{L + 1})}{C R_{\tilde{i}} (1-\alpha_{\tilde{i}})}\times( 1-\alpha_{\tilde{i}}^{L +1})\hspace{-0.1cm}\left[{ \int_{\hat{v}_{\tilde{i}}}^{\hat{v}_{{\tilde{i}}+1}}{\hspace{-0.5cm}v f_v(v) \mathrm{d}v}} - \hat{v}_{i}(F_v(\hat{v}_{\tilde{i}+1})-F_v(\hat{v}_{\tilde{i}}) )\right].\label{eq:reduction5}
\end{eqnarray}

\noindent Finally, \eqref{eq:reduction6} follows because $\int_{a}^{b}{x f(x) dx}\ge\int_{a}^{b}{a f(x) dx}$ if $a\ge 0$.

\begin{eqnarray}\hspace{-0.3cm}\frac{\partial {WT}}{\partial \hat{v}_{\tilde{i}}}h^2 \hspace{-0.3cm}&\ge& \hspace{-0.3cm}\frac{\mathbb{E}[b(\mathbf{s}_p)] f_v(\hat{v}_{\tilde{i}})(1-\alpha_{\tilde{i}}^{L + 1})}{C R_{\tilde{i}} (1-\alpha_{\tilde{i}})}\hspace{-0.1cm}\times\hspace{-0.1cm}( 1-\alpha_{\tilde{i}}^{L+1})\hspace{-0.1cm}\left[{ \hat{v}_{\tilde{i}}\hspace{-0.1cm}\int_{\hat{v}_{\tilde{i}}}^{\hat{v}_{{\tilde{i}}+1}}{\hspace{-0.5cm} f_v(v) \mathrm{d}v}} - \hat{v}_{i}(F_v(\hat{v}_{\tilde{i}+1})-F_v(\hat{v}_{\tilde{i}}) )\right]\label{eq:reduction6}
\\&=&0.\nonumber
\end{eqnarray}

\vspace{-0.2cm}

\noindent Thus, it follows that $\partial {WT}/\partial \hat{v}_{\tilde{i}}\ge 0$.

We prove part 2 of Lemma 1 by investigating the terms of the gradient $\partial {WT}/\partial \hat{v}_i = (h\partial g/\partial \hat{v}_i - g\partial h/\partial \hat{v}_i)/h^2$. We have $h\ge0$ and $\partial g/\partial \hat{v}_i\le0$ unconditionally. Furthermore, $\partial h/\partial \hat{v}_i\ge0~\forall i\ne \tilde{i}$ and $g\ge0$. Thus, $\partial {WT}/\partial \hat{v}_i\le0~\forall i\ne \tilde{i}$.

\vspace{-0.4cm}

\section{Proof of Lemma 3}

\noindent \textbf{Lemma 3. } \emph{Define $\mathcal{I} = \{\mathrm{argmax}~t_i\}$. If $\{\hat{v}_{i}; i\in\mathcal{I}$ or $i-1\in\mathcal{I}\}$ are jointly scaled to keep $\mathcal{I}$ fixed, then}

\begin{enumerate}
\item $\partial {WT}/\partial \hat{v}_{i}\ge0$ if $i\in\mathcal{I}$ and $i-1\not\in\mathcal{I}$
\item $\partial {WT}/\partial \hat{v}_i\le0$ if $i\not\in\mathcal{I}$ and $i-1\in\mathcal{I}$
\end{enumerate}

\textbf{Proof:} The special case of $|\mathcal{I}|=1$ is proved in Lemma 1. The case of $|\mathcal{I}|>1$ where the elements of $|\mathcal{I}|$ are non-consecutive also directly follows from Lemma 1 as one could jointly decrease $\{\hat{v}_i\}~\forall i\in\mathcal{I}$ and increase $\{\hat{v}_{i+1}\}~\forall i\in\mathcal{I}$ such that the set $\mathcal{I}$ is fixed. For the general case where some elements of $\mathcal{I}$ are consecutive, the set $\mathcal{I}$ can be divided into subsets of consecutive streams. For example, if $\mathcal{I}=\{1,3,4\}$, the first subset is $\{1\}$ and the second subset is \{3,4\}. Within each subset, $\partial {WT}/\partial \hat{v}_{i}\ge0$ for the lower-most stream satisfying $i\in\mathcal{I}$ and $i-1\not\in\mathcal{I}$ by part 1 of Lemma 1 and $\partial {WT}/\partial \hat{v}_{i}\le0$ for the upper-most stream satisfying $i\not\in\mathcal{I}$ and $i-1\in\mathcal{I}$ by part 2 of Lemma 1. Thus, there exist an infinitesimal step $\boldsymbol{\epsilon}=\{\epsilon_1,\cdots,\epsilon_S\}$ such that $\epsilon_i\ge0$ if $i\in\mathcal{I}$ and $i-1\not\in\mathcal{I}$ and $\epsilon_i\le0$ if $i\not\in\mathcal{I}$ and $i-1\in\mathcal{I}$ keeping $\mathcal{I}$ fixed and improving the objective and the result follows.

\section{Proof of Theorem 1}

\noindent \textbf{Theorem 1. } \emph{\textbf{Thresholding Policy:} The optimal loss visibility thresholds $\hat{\mathbf{v}}^* = \{\hat{v}_{i}^*\}_{i=2}^{S}$ satisfy}

\begin{equation}
F_v(\hat{v}_{i+1}^*) - F_v(\hat{v}_i^*) = \frac{R_{i}/r_i }{\sum_{j=1}^{S}{ R_j/r_j}}~\forall i = 1,\cdots,S
\end{equation}

\noindent \emph{where $r_i = (1-\alpha_i^{L + 1})/(1-\alpha_i)$.}

\textbf{Proof:} We present a convergent method that takes as input any feasible solution and obtains a solution with an improved objective satisfying the condition stated above. Start with any feasible solution and define the initial set of streams with the longest average transmission time $\mathcal{I} = \{i~\mathrm{s.t.}~t_{i} = \max_j t_j\}$. Construct an infinitesimal step $\boldsymbol{\epsilon}=\{\epsilon_1,\cdots,\epsilon_S\}$ such that $\epsilon_i\ge0$ if $i\in\mathcal{I}$ and $i-1\not\in\mathcal{I}$ and $\epsilon_i\le0$ if $i\not\in\mathcal{I}$ and $i-1\in\mathcal{I}$. By Lemma 3, there exist such an $\boldsymbol{\epsilon}$ such that $\mathcal{I}$ is unchanged and $WT(\hat{\mathbf{v}}+\boldsymbol{\epsilon})>WT(\hat{\mathbf{v}})$. Repeat until $\min_{i\in\mathcal{I},j\not\in\mathcal{I}}\{ t_i - t_j\} < \delta$ where $\delta$ is an arbitrarily small positive number. This necessarily increases $\mathcal{I}$. Repopulate $\mathcal{I}$ according to the new $\{\hat{v}_{i}\}$. Repeat until $\mathcal{I} = \{2,\cdots,S\}$. Thus, the optimal policy necessarily satisfies $t_1=t_2=\cdots=t_S$, equivalently, $(F_v(\hat{v}_{i+1})-F_v(\hat{v}_{i}))/(R_{i} (1-\alpha_i)) = (F_v(\hat{v}_{2})-F_v(\hat{v}_{1}))/(R_{1} (1-\alpha_{1}))~\forall i$. By taking $1 = \sum_i{F_v(\hat{v}_{i+1}) - F_v(\hat{v}_{i})}$, the Theorem follows.

\vspace{-0.1cm}
\bibliographystyle{ieeetr}
\bibliography{references}

\end{document}